\newif\iffigs\figsfalse
\figstrue

\input harvmac

\font\blackboard=msbm10 \font\blackboards=msbm7
\font\blackboardss=msbm5
\newfam\black
\textfont\black=\blackboard
\scriptfont\black=\blackboards
\scriptscriptfont\black=\blackboardss
\def\blackb#1{{\fam\black\relax#1}}

\def\BC{{\blackb C}} 
\def\BR{{\blackb R}} 
\def\BZ{{\blackb Z}} 
\def\BP{{\blackb P}}


\iffigs
  \input epsf
\else
  \message{No figures will be included. See TeX file for more
information.}
\fi

\Title{\vbox{\hbox{CLNS 96/1456}\hbox{UTTG--21--96}\hbox{CU-TP-813}
\hbox{\tt hep-th/9702030}\vskip -.5in }}
{Some Features of (0,2) Moduli Space}
\centerline{Ti-Ming Chiang$^\star$, Jacques Distler$^\dagger$ and Brian R.
Greene$^\ddagger$}
\vskip.3in
\centerline{\hbox{
\vtop{\hsize=1in\it
\centerline{$^\star$F.R. Newman Laboratory}
\centerline{of Nuclear Studies}
\centerline{Cornell University}
\centerline{Ithaca, NY 14853}
}\hskip .125in
\vtop{\hsize=3in\it
\centerline{$^\dagger$Theory Group}
\centerline{Department of Physics}
\centerline{University of Texas}
\centerline{Austin, TX 78712}
}\hskip .125in
\vtop{\hsize=1.25in\it
\centerline{$^\ddagger$Department of Physics and Mathematics}
\centerline{Columbia University}
\centerline{New York, NY 10027}
}
}}

{\parindent=-5pt

\footnote{}{\par
${}^\star$Email: {\tt chi@hepth.cornell.edu}\ .\par
${}^\dagger$Email: {\tt distler@golem.ph.utexas.edu}\ .\par
${}^\ddagger$Email: {\tt greene@shire.math.columbia.edu}\ .\par
}
}

\vskip .3in

\noblackbox

\def\WCP#1#2{\hbox{$\hbox{W\BC \BP}^{#1}_{#2}$}}

\def \LG{Landau-Ginzburg}

\def\CP#1{\hbox{$\hbox{\BC \BP}^{#1}$}}

\def\qt{\tilde{q}}

\def\td#1{\tilde{#1}}
\def\tilde{\widetilde}

\def\cO{{\cal O}}
\def\tablerule{\noalign{\hrule}}

\def\longra{\longrightarrow}
\def\vphone{\vphantom{\Gamma \over p}}
\def\vphtwo{\vphantom{{\td{E}^{'(j)}_1}}}
\def\vphthr{\vphantom{E_1^{(j)} \over \chi}}

We discuss some aspects of perturbative $(0,2)$ Calabi-Yau moduli space.
In particular, we show how models with different $(0,2)$ data
can meet along various sub-loci in their moduli space. In the
simplest examples, the models differ by the choice of desingularization
of a holomorphic V-bundle over the same resolved Calabi-Yau base while
in more complicated examples, even the smooth Calabi-Yau base manifolds
can be topologically distinct. These latter examples extend and
clarify a previous observation which was limited to singular Calabi-Yau
spaces and seem to indicate a multicritical structure in moduli space.
This should have a natural F-theory counterpart in terms of the moduli
space of Calabi-Yau four-folds.

\Date{2/97}

\newsec{Introduction}

\lref\rWittennewissues{E. Witten, ``New issues in manifolds of SU(3) holonomy''
, {\it Nucl. Phys.} {\bf B268} (1986) 79}
\lref\rWphases{E. Witten, ``Phases of $N=2$ theories in two dimensions'',
{\it Nucl. Phys.} {\bf B403} (1993) 159, {\tt hep-th/9301042}}
\lref\rnrefs{R. Friedman, J. Morgan and E. Witten, ``Vector Bundles and
F Theory'', {\tt hep-th/9701162} and references therein; M. Bershadsky,
A. Johansen, T. Pantev and V. Sadov, ``Four-Dimensional Compactifications
of F-theory'', {\tt hep-th/9701165} and references therein}
\lref\rSW{E. Silverstein and E. Witten, ``Criteria for conformal invariance of
(0,2) models'', {\it Nucl. Phys.} {\bf B444} (1995) 161, 
{\tt hep-th/9503212} }
\lref\rAG{P. Aspinwall and B. Greene, ``On the Geometric
Interpretation of N = 2 Superconformal Theories'', {\it Nucl. Phys.}
{\bf B437} (1995) 205, {\tt hep-th/9409110}}
\lref\rMP{D. Morrison and M. Plesser, ``Summing the instantons: Quantum 
cohomology and mirror symmetry in toric varieties'', {\it Nucl. Phys.}
{\bf B440} (1995) 279, {\tt hep-th/9412236}}
\lref\rmirrortwo{R. Blumenhagen, R. Schimmrigk and A. Wisskirchen, ``(0,2)
Mirror Symmetry'', {\tt hep-th/9609167}; 
R. Blumenhagen and S. Sethi, ``On Orbifolds
of (0,2) Models'', {\tt hep-th/9611172}}
\lref\rDGM{J. Distler, B. Greene and D. Morrison, ``Resolving 
Singularities in (0,2) Models'', {\it Nucl. Phys.} {\bf B481} (1996) 289, 
{\tt hep-th/9605222}}
\lref\rDK{J. Distler and S. Kachru, ``Duality of (0,2) String Vacua'',
{\it Nucl. Phys.} {\bf B442} (1995) 64, {\tt hep-th/9501111}}
\lref\DKthree{J. Distler and S. Kachru, ``Quantum Symmetries and
Stringy Instantons," Phys. Lett. {\bf B336} (1994) 368, {\tt
hep-th/9406091}}
\lref\rDist{J. Distler, ``Notes on (0,2) Superconformal Field Theories'',
in Trieste HEP Cosmology 1994, {\tt hep-th/9502012}}
\lref\rAGM{P.S. Aspinwall, B.R. Greene and D.R. Morrison, ``Calabi-Yau
Moduli Space, Mirror Manifolds and Spacetime Topology Change in String
Theory'', {\it Nucl Phys.} {\bf B416} (1994) 414, {\tt hep-th/9309097}}
\lref\rKSS{S. Kachru, N. Seiberg and E. Silverstein, ``SUSY Gauge
Dynamics and Singularities of 4d N=1 String Vacua'', {\it Nucl. Phys.}
{\bf B480} (1996) 170, {\tt hep-th/9605036}}
\lref\rWinst{E. Witten, ``Small Instantons in String Theory'',
{\it Nucl. Phys.} {\bf B460} (1996) 541, {\tt hep-th/9511030}}
\lref\rGMV{B. R. Greene, D. R. Morrison and C. Vafa, ``A Geometric
Realization of Confinement'', {\it Nucl. Phys.} {\bf B481} (1996) 513,
{\tt hep-th/9608039}}

Recent developments have yielded tremendous insights into numerous
areas of string theory. The physics of
left-right symmetric Calabi-Yau string compactifications is one
such area in which our understanding has been sharply refined
through the discovery of a variety of remarkable physical phenomenon.
As has long been recognized, these $(2,2)$ string models are but
a small subclass of the more general $(0,2)$ models initially introduced
in \rWittennewissues. After a number of years of limited progress,
the work of \rWphases\ has given rise to renewed hope that our understanding
of $(0,2)$ Calabi-Yau string models may one day be on par with that
for $(2,2)$ models. This hope has received support from recent developments
in string duality, most notably F-theory. In particular, F-theory on a special
class of Calabi-Yau four-folds is dual to heterotic $(0,2)$ vacua 
on Calabi-Yau three-folds, with the left-right symmetric geometric
moduli of the former capturing the left-right asymmetric moduli of the latter.
Fully exploiting this tool, though requires a more precise dictionary between
the two than has presently been achieved, so we shall perform our
analysis directly in the $(0,2)$ heterotic models themselves.  It would be
of great interest to deepen our explicit understanding of this dictionary in
order to be able to approach the kind of questions we study here in the
context of four-folds \foot{Very recently, a number of new results 
\rnrefs\ have sharpened the F-theory/heterotic dictionary. We have not 
determined, as yet, how the work presented here relates to those results.}.

The present paper is concerned with a few unusual features of the moduli
space of $(0,2)$ Calabi-Yau string compactifications. Recall that
a perturbative $(0,2)$ model is specified by a choice of a Calabi-Yau
manifold $M$ and a holomorphic vector bundle $V$
satisfying
\eqn\efcond{c_1(V) = 0}
and
\eqn\escond{c_2(V) = c_2(T)}
where $T$ is the holomorphic tangent bundle of $M$
\foot{Generally $c_1(V)$ is only required to
vanish mod 2 to admit spinors. This additional freedom has
not as yet been explored in any detail.}.
One immediate solution to these conditions is
$V = T$ and this yields the familiar $(2,2)$ case.
Our interest is on solutions which are not of this
special form. Progress in studying these more general
solutions was hampered by the substantial complexity
and relative paucity of known examples. Witten's
linear sigma model provides a powerful tool
for straightforwardly generating new solutions.
As we will briefly review in section II, by working at
the level of two-dimensional non-conformal quantum field
theories, the linear sigma model dispenses with much
of the complexity of the associated conformal fixed
point theory, while retaining many of its essential
features. Furthermore, models derived from the linear
sigma model approach, as argued in \rSW\
are particularly robust as they avoid instanton
destabilization. For these reasons, all of our remarks
will  be confined to $(0,2)$ models with a linear
sigma model realization.

In \rWphases, the global
discussion of both $(2,2)$ and $(0,2)$ models
in an irreducible toric variety such as
a hypersurface in a 
weighted projective space was limited, for the most part,
to a one-dimensional slice through the K\"ahler
moduli space. Typically, this slice probes a family
of singular Calabi-Yau spaces.
 Through the link between the physics of
linear sigma models and the mathematics of toric
geometry, it was shown in \refs{\rAG,\rMP}\
how to systematically include the other K\"ahler moduli
in the context of $(2,2)$ models
and thereby gain access to the full K\"ahler moduli space
\foot{We note that this approach actually gives access to
the full toric part of the K\"ahler moduli space, which is
sometimes a proper subspace of the full K\"ahler moduli space.}.
As shown through local analysis in \rWphases\
and global analysis in \rAGM, inclusion of these other degrees
of freedom in $(2,2)$ models yields important physical phenomenon such as
topology changing flop transitions in left-right symmetric
compactifications. With the importance of studying the
properties of the full K\"ahler moduli space of
$(2,2)$ models thereby established, a natural direction to
pursue is a similar study of $(0,2)$ models. This is a more
delicate question since the extra modes included must
not only resolve singularities in the Calabi-Yau base $M$,
but they must also repair singularities in the vector bundle
$V$. In \rDGM, this problem was studied, and a fairly
systematic solution was offered. The new features of
bundle singularities was addressed in the context
of the linear sigma model.
Amongst the most striking new local features encountered
are examples in which the nonlinear sigma model gauge data is naturally
described in the language of sheaves,
 with corresponding ``protuberance''
structures in the associated linear sigma model.
In the present paper our interest is in the more global aspect
of these $(0,2)$ models.

In section II we review the method of \rDGM\ for
resolving a class of the perturbative singularities which
can arise in $(0,2)$ models. We frame this approach by discussing,
 in section III,  the kinds
of singularities --- both perturbative and nonperturbative ---
which can arise in the $(0,2)$ setting, and contrast the situation
with that relevant for $(2,2)$ models. In section IV
we focus on a variety of examples which
serve to illustrate some novel features of resolved
$(0,2)$ moduli space. We discuss, for instance, examples in which
there are multiple resolutions of given singular
geometric data. Our prime focus, however, is on carrying
forth an observation made in \rDK\ regarding the possibility of a
novel  duality between various $(0,2)$ models.
These original observations were left somewhat inconclusive
as only models along singular loci in moduli space were
discussed. Applying the desingularizing methodology of
\rDGM\ to these examples proves to be a bit subtle;
however we are able to construct new examples which
avoid these complexities. In this way we are able to
clarify and extend our understanding of these striking
dual examples. In section V we briefly describe the
constraints which stability seems to require, and finally
we offer some conclusions in section VII.

\newsec{Constructing Resolved $(0,2)$ Models}

In this section we briefly review the algorithm put
forth in \rDGM\ for desingularizing  a particular
class of $(0,2)$ models.

As mentioned above, the data for a $(0,2)$ model consists
of a Calabi-Yau base $M$ and a vector bundle $V$. The
typical class of Calabi-Yau spaces dealt with in the literature
over the years are complete intersections in toric varieties.
Of these, the most familiar are those in which the initial
toric variety is a weighted projective space, or possibly
a product of such spaces. These are the ones we shall consider.
The essential point is that weighted projective spaces in which
the individual weights are not all relatively prime have
singularities from the projective identifications. The full
parameter space of the corresponding physical model is found
only by resolving these singularities.
Bundles ---
or more precisely V-bundles --- over the original singular
space must correspondingly be augmented by data specifying their
behaviour over regions of the base which are modified by the
desingularizing procedure. All the while, attention must
be paid to avoid spoiling any of the delicate geometric and
topological consistency requirements to have a genuine solution.

To accomplish this task we use the formalism of Witten's linear
sigma model augmented by  methods from toric geometry.
To keep the notation from getting out of hand, we
consider a Calabi-Yau hypersurface $M$ in a single
weighted projective space, $\WCP{n-1}{q_1, q_2, ..., q_n}$. The
procedure for defining a $(0,2)$ model on  $M$ is as
follows:

One considers right moving
$(0,2)$ chiral superfields $P$ (with scalar and fermionic components $p$ and
$\pi$) and $\Phi_1, ... , \Phi_n$ (respectively with scalar
and fermionic  components $\phi_i$ and
$\psi_i$) with U(1) gauge charges $q_0, q_1, ... , q_n$, as well as
left moving $(0,2)$ fermi multiplets $\Gamma$ (with first component $\gamma$)
and $\Lambda_1, ... , \Lambda_m$ (respectively with first component
$\lambda_i$) with charges $\qt_0, \qt_1, ... , \qt_m$.

	The superpotential then takes the following form:
\eqn\eSP{\int d^2 z d\theta (\Gamma G(\Phi_i) + \sum_{a} \Lambda^a P
F_a(\Phi_i))}
where $G$ and $F_a$ ($a = 1, ... , m$) are homogeneous polynomials.
$U(1)$-invariance dictates that the degrees of $G$ and $F_a$ ($a = 1, ... , m$)
are respectively $-\qt_0$, $-\qt_1 - q_0, ... , -\qt_m - q_0$.

\subsec{Yukawa couplings and anomaly cancellation}
The  gauge charges of these fields are not fully arbitrary as they
are subject to anomaly cancellation conditions. To understand the
geometrical meaning of these conditions, we examine
the various Yukawa-couplings in the action.

	Let us first consider the right moving fermions. From $\eSP$ we
obtain the coupling:
\eqn\eyukawarf{\sum_{i} \gamma \psi_i {{\partial G} \over {\partial \phi_i}}}
This gives mass to a linear combination of the $\psi_i$'s, if the
Calabi-Yau is transverse in the sense that
${{\partial (\gamma G)} \over {\partial \phi_i}} d\phi_i$
\foot{This form easily generalizes to complete intersections.}
does not vanish on the manifold. Another coupling with the gaugino
$\alpha$ of the bosonic gauge symmetry:
\eqn\eyukawarg{-\sum_{i} q_{i} \alpha \psi_i \overline{\phi_i} - q_0 \alpha
\pi \overline{p}}
gives mass to a different linear combination of the $\psi_i$'s,
provided $p$ vanishes (as in a typical Calabi-Yau phase).
These two combine to imply that the massless right-moving fermions transform
as sections of the cohomology of the following sequence (``monad"):
\eqn\eseqtb{0 \longrightarrow \cO {\buildrel \otimes q_{i} \phi_i \over
\longrightarrow} \bigoplus_{i=1}^{n} \cO(q_i)
{\buildrel {\otimes {{\partial G} \over {\partial \phi_i}}} \over
\longrightarrow}
\cO(-\qt_0) \longrightarrow 0}

In simple terms, this means that they transform as sections of the tangent
bundle $T$. As the Calabi-Yau condition requires that $c_1(T) = 0$,
we obtain the following condition on the charges:
\eqn\eanomrc{\sum_{i=1}^{n} q_{i} = - \qt_{0}}

The analysis for the left-moving fermions follows a similar sequence of steps.
{}From the superpotential we obtain the coupling
\eqn\eyukawalf{\sum_{a} \pi \lambda^{a} F_{a}}
This gives mass to a certain combination of $\lambda^{a}$'s provided that
a similar transversality condition holds, i.e. that the $F_{a}$'s do not
simultaneously vanish
anywhere on the manifold. \eyukawalf\ is the analog of \eyukawarf\ in the
left-moving case.
To see a similar analog for $\eyukawarg$,
we recall that in a $(2,2)$ model, the gauge multiplet may be decomposed as
a combination of a bosonic and fermionic $(0,2)$ gauge multiplet.
Our interest being in $(0,2)$ models which are generally not holomorphic
deformations of their $(2,2)$ cousins, we are led to a notion of {\it fermionic
gauge} symmetries that can be incorporated  independently of the
bosonic symmetries. The
introduction of $l$ such symmetries \rDist\
requires the existence of $l(m+1)$ homogeneous polynomials
$E^a_j$ , $a = 0, ... , m$, $j = 1, ... , l$ which form part of the action on
the fermi multiplets:
\eqn\efermsym{\Gamma \rightarrow \Gamma + 2 P E^0_{j}(\Phi) \Omega^{j} \qquad
\Lambda^{a} \rightarrow \Lambda^{a} + 2 E^a_{j}(\Phi) \Omega^{j}}
where the $\Omega^{j}$'s  are $U(1)$-neutral fermionic chiral multiplets.
Often times, the fermionic symmetries are
chosen in order to obtain a sensible \LG\ interpretation, so that,
for instance, there are no chargeless fermions  in the \LG\ phase.
Invariance of \eSP\ under \efermsym\ then implies
\eqn\esuperinv{E^0_j G + \sum_{a = 1}^{m} E^a_j F_a = 0}

Assuming we can find such $E$'s, we still have to ensure the
invariance of the left-moving fermion kinetic terms under
\efermsym. To do so,
we introduce $l$ additional {\it unconstrained} $U(1)$ neutral
fermionic
superfields $\Sigma^j$ transforming as
$\Sigma^j \rightarrow \Sigma^j + \Omega^j$ and couple them appropriately
to the other fields in the action. For our purposes, it is essential to know
only the two main consequences of doing this. First of all, a scalar potential
of the following form arises:
\eqn\efermgpot{\sum_{j} \sum_{k} (|p|^2 \overline{E^0_j(\phi_i) \sigma_j} 
E^0_k(\phi_i) \sigma_k + \sum_{a=1}^{m} \overline{E^a_j(\phi_i) \sigma_j}
E^a_k(\phi_i) \sigma_k)}
where $\sigma_j$ is related to the gauge invariant quantity
$\overline{D}_{+} \Sigma^j$ by
\eqn\edplussigma{\overline{D}_{+} \Sigma_j = {1 \over 2} (\sigma_j + \theta^{-}
\beta_j + \theta^{-} \theta^{+} \partial_{\overline{z}} \sigma_j)}
Secondly, the $\Sigma_j$ couplings also contain a Yukawa coupling of the form
\eqn\eyukawals{\sum_{j} \overline{\beta}_{j} (\overline{p E^0_j(\phi_i)} \gamma
+ \sum_{a=1}^{m} \overline{E^a_j(\phi_i)} \lambda_a)}
which, in a typical Calabi-Yau phase where $p$ has no expectation value,
gives mass to $l$ combinations of the $\lambda$'s, provided
that $(E^1_j, ... , E^m_j)$ form $l$ linearly-independent vectors at
all points on the manifold. This gives us the desired analog of
\eyukawarg. Combined with \eyukawalf, we therefore
see that the massless left-moving fermions transform as sections of
the bundle $V$ over $M$ defined by the cohomology of the following sequence:
\eqn\eseqtz{0 \longra \oplus^l \cO {\buildrel {\otimes E^a_j} \over \longra}
\oplus_{a=1}^{m} \cO(\qt_a) {\buildrel {\otimes F^a} \over \longra}
\cO(-q_0) \longra 0}

Equations \efcond\ and \escond\ then gives us the remaining conditions on
the charges\foot{The second condition is somewhat stronger than just requiring
that the Chern classes be equal since it amounts to ignoring relations
between the ambient projective cohomology classes restricted to $M$. It also
however follows {}from cancelling the worldsheet gauge $U(1)$ anomaly.}.
\eqn\efccond{q_0 = -\sum_{a=1}^{m} \qt_a}
\eqn\esccond{\sum_{i=1}^{n} q_i^2 = \sum_{a=1}^{m} \qt_a^2}

{}From the point of view of
field theory, \esccond\ is the condition that $U(1)$ gauge symmetry is
nonanomalous. \eanomrc\ and \efccond\ are the conditions that ensure the
existence of a nonanomalous global $U(1)\times U(1)_R$ symmetry, which becomes
the left- and right-moving $U(1)$s in the conformal limit.
Before generalizing these conditions as appropriate for a {\it nonsingular}
base, we briefly review the basic features of
the 1-dimensional K\"ahler moduli space.

\subsec{The one-parameter moduli space}
Here the typical complete $(0,2)$ bosonic potential is:
\eqn\ebosonicp{U = |G(\phi_i)|^2 + |p|^2 \sum_{a} |F_{a}|^2 + {1 \over 2e^2}
D^2 + |\sigma|^2 (|p|^2 |E^0(\phi_i)|^2 +
\sum_{a} |E^a(\phi_i)|^2)}
where
\eqn\eDterm{D = -e^2(\sum_{i=1}^{n} q_{i} |\phi_i|^2 + q_0 |p|^2 - r)}

Now the $q_{i}$'s are positive since they are the weights of the the ambient
projective space. Furthermore the $\qt_{a}$'s for $a = 1, .. , m$ are positive
because otherwise the bundle over the (possibly singular) base as defined
in \eseqtz\ would not be stable. Thus from \efccond\ we obtain that $q_0$ is
negative. The semiclassical
vacuum is found by setting $U = 0$ which implies
that $D$ and $G$  vanish and that either $p$ or all the $F$'s must
vanish. Let us  analyze this in detail, first for the case when $r > 0$.

	In this case, $D$ vanishing means not all the $\phi$'s can vanish.
Hence as $G$ vanishes and one of our assumptions was that
the $F$'s do not simultaneously vanish at any point in the manifold,
$p$ must also vanish.
Furthermore we also assumed that the $E^a$'s for
$a = 1, ... , m$ do not simultaneously vanish when $G$ does, which means
{}from \ebosonicp\ that the $\sigma$ must vanish.
Looking at the D-term
again then gives $\sum_i q_i |\phi_i|^2 = r$, which, after
$U(1)$ identifications,
is just a copy of $\WCP{n-1}{q_1, ..., q_n}$. The locus $G = 0$ then defines
a Calabi-Yau in this weighted projective  space. Actually, since $p$ transforms
nontrivially under the gauge group it is natural to consider the manifold
as embedded in the total space of $\cO(q_0)$ over $\WCP{n-1}{q_1, ..., q_n}$.
The latter space is nothing but the toric variety found by setting
$D$ to zero modulo gauge equivalences without any further constraints.
The resulting model
is therefore the Calabi-Yau sigma model on $M$ with the left-moving
fermions transforming as sections of $V$.

One apparent problem with the previous analysis is that,
unlike in $(2,2)$ models, there is no
relation between $q_0$ and $q_1, ... , q_n$. The charge relations
\eanomrc\ and \efccond\ relate $q$'s and $\qt$'s instead. This would contribute
to a 1-loop renormalization of $r$ proportional to $q_{tot} = \sum_{i=0}^{n}
q_i$ so that the theory would be strongly renormalized from our semiclassical
treatment above. A means of avoiding this
was described in \rDist\ and amounts
to introducing a pair of chiral superfields, termed ``spectator'' superfields,
 $S$ and $\Xi$ with charges
$-q_{tot}$ and $q_{tot}$ respectively. They contribute
to the superpotential a term of the form $\Xi S$, which means that they are
massive and can be set to zero in determining the classical vacuum at any
RG-invariant radius. However, their 1-loop contribution cancels the
renormalization of $r$. Effectively, the linear sigma model trades the two
parameters (the coefficient of $\Xi S$ in the superpotential and $r+i\theta$)
for a single RG-invariant parameter, which we will call $r+i\theta$. In
particular, our previous analysis remains valid with the addition of such
spectators.

Having thus discussed the scenario for $r > 0$, we turn to the
other interesting
case of $r < 0$. Here, $D = 0$ implies that $p$ cannot vanish, and
 hence
that the $F$'s must simultaneously vanish, which occurs, as $G$ vanishes
as well, only when all the $\phi_i$'s are zero.
This means that $|p|$ is
$\sqrt{r/q_0}$. Our vacuum is thus, after $U(1)$ gauge identifications, a
single point with $(\phi_1, ... , \phi_n, p) = (0, ..., 0, \sqrt{r/q_0})$.
If $E^0(0, ..., 0)$ does not vanish at this point, $\sigma$ again
drops out in the infrared. Actually \rWphases\ there is a residual
$\BZ_{-q_0}$ gauge symmetry in the vacuum, which is embedded naturally in
$\BC^{n} / \BZ_{-q_0}$. We thus obtain an orbifold of an
\LG\  model defined over the vacuum.

The Calabi-Yau sigma model and \LG\ orbifold regimes described
above form only a one-dimensional slice of the
full K\"ahler moduli space of the theory. The other phases are found by
including $N=2$ superconformal marginal operators that can be interpreted
as modes that geometrically resolve
singularities in the base manifold. Most of these modes are captured
naturally in the formalism
of toric geometry.
A detailed introduction to toric geometry for physicists can be found
in \rAGM; here we will freely make use of this formalism.
Toric geometry allows our Calabi-Yau to be associated with a
Gorenstein cone in $\BR^{d}$ with $k$ edges. This is reflected in a
new linear sigma model with $s = h^{1,1} - 1 = k - d$ extra chiral superfields
and $U(1)^{s+1}$ gauge symmetry where $h^{1,1}$ is the hodge number of the
resolved space \foot{More precisely $s = h^{1,1}_{toric} - 1$ where
$h^{1,1}_{toric}$ is the number of divisors that arise from the toric
ambient space.}. The charges of the $k = s + d$ chiral superfields
are determined by the kernel of the $(k + 1) \times d$ point set matrix
formed from the vertices and interior of the polytopic base of the cone \rAG.
Corresponding to each regular triangulation of the
base is a phase of the linear $\sigma$-model defined by a region in the
parameter space of the $s + 1$ Fayet-Iliopoulos coefficients $r_i$.

\subsec{Desingularization of $(0,2)$ models}
	The $(2,2)$  toric formalism thus naturally allows for
desingularization
of the base, and along with it
the desingularization of the tangent bundle.
In the $(0,2)$ case however we must supply more information to pull
the original $V$-bundle $V $ back to the nonsingular base.
In \rDGM\ a straightforward approach for doing this was proposed: if
the resolved base has a $h^{1,1}$ dimensional K\"ahler moduli space,
then line bundles over it --- which can be classified by their
first Chern class --- are specified by the equivalent data
of $h^{1,1}$ integers: their
charges under each of the $h^{1,1}$ $U(1)$ gauge factors.
These charges must be such as to meet conditions \efcond\ and \escond,
which are physically equivalent to the cancellation of all anomalies
in the linear sigma model.
Thus, the desingularization
 of the original $V$-bundle is obtained by searching for
$h^{1,1} - 1$ additional charges for each line bundle in the definition
of the gauge bundle meeting the anomaly cancellation conditions.
It would be nice to have a more intrinsic procedure, but this
is our present formulation.

	So to summarize, the steps needed in defining a $(0,2)$ model on a
hypersurface $M$ in a weighted projective space are as follows:

{$\bullet$ Construct the appropriate toric data for the base, giving us
the the charges of the left-moving fermion $\Gamma$ and $n$ right-moving
bosons $\phi_i$ under a $U(1)^{s}$ gauge group. The set of charges can be
denoted by $\qt_0^{(k)}$ and $q_{i}^{(k)}$ where $i$ runs from $1$ to $n$ and
$k$ from $1$ to $s$.}

{$\bullet$ Construct an appropriate set of charges $q_0^{(k)}$ and
$\qt_{a}^{(k)}$ where $a$ runs from $1$ to $m$ satisfying the
following anomaly cancellation conditions which are a generalization of
\efccond\ and \esccond:
\eqn\eanomfc{\sum_{a=1}^{m} \qt_{a}^{(k)} + q_0^{(k)}= 0}
\eqn\eanomsc{\sum_{a=1}^{m} \qt_{a}^{(j)} \qt_{a}^{(k)} + \qt_0^{(j)}
\qt_0^{(k)} = \sum_{i=1}^{n} q_{i}^{(j)} q_{i}^{(k)} + q_0^{(j)} q_0^{(k)}}
where $j$ and $k$ run from $1$ to $s$. The charges $q_0^{(k)}$ is to be
interpreted as that of a right-moving bosonic superfield $P$ and
$\qt_{a}^{(k)}$
are the charges of $m$ left-moving fermionic superfields $\Lambda^a$.}

{$\bullet$ Choose a number $l$ (which can be 0) of fermionic gauge symmetries,
and determine suitable $E$'s and $F$'s satisfying \esuperinv. The massless
left-moving
fermions thus transform under a vector bundle $V$ which is defined by
the following two {\it exact} sequences (as a generalization of
\eseqtz):
\eqn\eSeqtzf{ 0 \longra \oplus^{l} \cO {\buildrel {\otimes E^a_j} \over
\longra}
\oplus_{a=1}^{m} \cO(\qt_a^{(1)}, \qt_a^{(2)}, ... , \qt_a^{(s)})
 {\longra} {{\cal E}} \longra 0}
\eqn\eSeqtzs{ 0 \longra V \longra {{\cal E}} {\buildrel
{\otimes F_a}
\over \longra}
\cO(-q_0^{(1)}, -q_0^{(2)}, ... , -q_0^{(s)}) \longra 0}
}

The maps above are as follows:
\eqn\emapE{\otimes E^a_j : (c_1, ... , c_l) \rightarrow (\sum_{j=1}^{l} c_j
E^1_j (\phi_i), ... , \sum_{j=1}^{l} c_j E^m_j(\phi_i))}
\eqn\emapF{\otimes F_a : (\xi_1(\phi_i), ... , \xi_m(\phi_i)) \rightarrow
\sum_{a=1}^{m} \xi_a(\phi_i) F_a(\phi_i)}

\newsec{Singularities in $(0,2)$ Models}

Singularities in a linear $\sigma$-model are generally
detected by searching for noncompact directions 
in the classical vacuum. The rationale for this follows, roughly
speaking, from the possibility of bad behaviour
in the path integrand used in calculating correlation functions
at large values of the
relevant {\it noncompact} fields, i.e. fields that have a noncompact
range in the $\sigma$-model vacuum. 
A better indication of  a singularity is a non-compact
direction in the full quantum effective potential \rWphases. 	
For the most part we shall ignore these complexities and base
our analysis on the classical potential; these aspects deserve further
study.

In $(2,2)$ models,  perturbative singularities are known to arise,
for a one-parameter case, when the complex Fayet-Iliopoulos 
parameter vanishes, classically,  (more precisely,
 when it is equal to  a particular nonzero
value, taking one-loop linear sigma model corrections into account)
\rWphases\ or when the manifold becomes singular. In the former case,
the noncompact field is the gauge field $\sigma$ while in the latter
it is a right-moving boson $p$. They correspond respectively to  
K\"ahler and complex structure degenerations.

For $(0,2)$ models, the story has some similarities as well as
some differences. In particular,
the presence or absence of fermionic gauge symmetries is a 
crucial factor.
In the $(2,2)$
case, if the bosonic gauge fields were absent in the scalar potential
\rWphases, there would be no
singularities in the K\"ahler moduli space of a smooth Calabi-Yau
$M$.
Similarly, in the $(0,2)$ case, if there are no fermionic gauge
symmetries,
singularities can only possibly occur when the bundle is
{\it nontransverse},
i.e. when the $F$'s simultaneously vanish at a point on the
``manifold''.
This means
that, for instance, there are a priori no physical singularities associated
with
a base space having bad {\it complex structure}, so long as none of
the bosonic fields acquire a noncompact range.
The lesson we learn
{}from this
is that the natural arena of $(0,2)$ models is geometrically realized in
terms of coherent sheaves over base spaces
that can be relatively badly behaved.

When fermionic gauge symmetries are present, a new potential difficulty
arises. Given a model defined as a monad over the singular manifold, when we
resolve the manifold, we may find that there is no choice of $E$'s  on the
{\it resolved} manifold, $M$, for which the first map in (2.19) is injective.
In that case, the resulting (0,2) theory is, classically,
{\it generically} singular, for
there is a noncompact branch to the vacuum, corresponding to some $\sigma_j$
having a flat potential.   It might be that these models require 
non-perturbative
effects to be taken into account everywhere in their moduli space, or
it might be that they are only physically sensible before resolving
the geometrical singularities. Although each possibility is quite interesting,
 in this paper, we will skirt 
the issue by eschewing such models.

When we have a \LG\ phase, one can sometimes show that fermionic gauge 
symmetries
are required for consistency. In the absence of a \LG\ phase (or a detailed 
check
of the stability of the bundle), we need to make some assumptions about what
fermionic gauge symmetries are present. We will assume that the only ones which
are present are those which lead {\it classically} to a 
generically-nonsingular (in the sense
that there are, generically, no noncompact directions in the vacuum) model on
the resolved manifold. We should point out that we presently lack
a first principles argument for which fermionic gauge
symmetries a particular model requires; the assumption we make here is 
consistent
with our present level of understanding but deserves further investigation.

Having thus summarized what is to come, we begin with a
detailed discussion
by examining the bosonic
potential for a $(0,2)$ model based on a Calabi-Yau hypersurface $M$ in
$\WCP{n-1}{q_1,...,q_n}$. Similar conclusions can be obtained for other models
based on more general complete intersections in toric varieties.

\subsec{Compactness of p}
Let us first determine when, in the notation of the previous section, any of
the $\phi$'s can become noncompact. This is for our purposes actually
equivalent to
determining when $p$ can become noncompact, because in all the examples
described
in this paper, there is always a special basis of the gauge group $U(1)^s$
under
which all the $\phi$'s have nonnegative multi-charges, while
 $p$ and $\gamma$ have negative
multi-charges. This is certainly
true for most $(2,2)$ models, for instance.
Equation \ebosonicp\ then clearly shows
that if some $\phi_i$ were to grow arbitrarily while maintaining $U = 0$ at a
fixed point in the K\"ahler moduli space, $p$ must also grow with $\phi_i$ to
maintain the vanishing of the $D$-terms. Thus we need only consider when $p$ or
the $\sigma_j$'s could become noncompact.

We examine first the case with $p$.  To be concrete, consider a {\it typical}
smooth Calabi-Yau phase. From
\ebosonicp\ we see
that $p$ can grow arbitrarily large only where $G$ as well as all the $F$'s
simultaneously vanish in the ambient toric variety.
Here the ambient toric variety is the total space of a negative line bundle
(which $p$ transforms under) over a compact toric variety $C$ in which the
Calabi-Yau $M$ is embedded. If the $F$'s simultaneously vanish at some
locus $S$ in $C$, then $p$ would become noncompact over $S \cap M$. If this
happens, we then obtain a singularity in the gauge bundle which is the
$(0,2)$ analog of being on the discriminant locus of a $(2,2)$ model.
Indeed, in \rSW\ it was shown that for
$(0,2)$ deformations of the quintic, the  correlation function of 3
${\bf 27}$'s develop a simple pole at such a singularity.
In the case where $M$ is a $K3$ and $S \cap M$
consists of a set of points, this has been understood from the
point of view of nonperturbative gauge symmetries in 6 dimensions \rWinst.

	While the discussion just given holds for a typical Calabi-Yau,
one may wonder if it leads to similar conclusions for all $(0,2)$ vacua
with an obvious geometrical interpretation.
The answer is no, and the reason is that the discussion
just given relied on $p$ being unconstrained by the $D$-terms
at all points of $C$. In some cases, however, as is familiar
{}from studies of certain $(2,2)$ models, the range of $p$
can be forced to be
compact over a toric subvariety $Z$ in $C$. If the $F$'s vanish
only on $Z$, then there
would be no singularity in the underlying linear $\sigma$-model as
$p$
is then compact over $M$. In the $(2,2)$ case, the relevant phases
where this
occurs are related to ordinary Calabi-Yau phases in the same
moduli space by a flop in a
{\it noncompact} direction \rGMV. In the $(0,2)$ case
the situation is similar, an example being given by page 19 of \rDGM.
In that example, the degrees of the $F^a$'s were
{\it not generic} enough
to allow for avoidance of gauge bundle singularities on the base.
Nonetheless, the $D$-terms constrain $p$ to be compact over the common
vanishing locus of the $F$'s.
We may
thus view the compactness of $p$ as the linear $\sigma$-model's
way of
coping with what would have been an ``unavoidable'' singularity.
$V$, as it turns out,
is then no longer a vector bundle but rather a coherent sheaf. 
We will see later that sheaves commonly occur as
perturbative
vacua in $(0,2)$ models. For at least some classes of examples
in which $p$ actually is non-compact, \rKSS\ have shown
that there is a non-perturbative mechanism leading to a sensible
physical theory. We will not consider such examples in this paper.

\subsec{Compactness of $\sigma_j$}
{}From 	\efermgpot,
we see that if the matrix
$M$ whose elements are given by $M_{jk} = |p|^2 \overline{E^0_j} E^0_k +
\sum_{a} \overline{E^{a}_{j}} 
E^{a}_{k}$
is not positive definite,
there will be a noncompact direction in the space
of $\sigma_j$'s.
This yields  a situation analogous to a K\"ahler degeneration in
a $(2,2)$ model, which occurs at non-generic points on the walls of a
K\"ahler cone in the enlarged K\"ahler moduli space. In the $(0,2)$
case, though, there are some differences. The essential difference
for our present discussion is that K\"ahler degenerations in 
$(2,2)$ models occur at complex codimension one in the  enlarged
K\"ahler moduli space, at those loci where the (2,2) linear sigma model gets 
an unbroken $U(1)$ gauge symmetry. More generally, in complex codimension $n$, 
the linear sigma model has an unbroken $U(1)^n$, and there are $n$ 
noncompact directions in the vacuum.

In $(0,2)$ models, on the other hand, the singularity loci in the moduli
space, associated to zero eigenvalues of the matrix $M$, do not directly 
track the unbroken gauge symmetry
points (as they do in the $(2,2)$ case). This means that the number
of zero eigenvalues (the number of noncompact $\sigma$-directions) is largely 
decoupled from the rank of the unbroken
gauge symmetry and can only be determined by direct calculation.

This is important. While we will continue to speak of the ``phases" of the 
(0,2) model, and the walls of the K\"ahler cone, at which some $U(1)$ gauge 
symmetry is restored, these are {\it not} necessarily the locations of 
singularities of the (0,2) model. {\it Those} are determined by the 
fermionic gauge symmetries and the associated coupling to $\sigma$. In the 
simplest examples -- those with no fermionic gauge symmetries -- there are no 
singularities in the K\"ahler moduli space.

One of the tricky matters in formulating a (0,2) linear sigma model is 
deciding when, and how many, fermionic gauge symmetries ought to be present. 
Sometimes the answer is forced upon you by consistency considerations. 
Clearly, if the bundle $V$ defined by \eSeqtzf\ and \eSeqtzs\ is stable for 
$l$ fermionic gauge symmetries, it will not be stable if we tried to 
formulate it with $l-1$ fermionic gauge symmetries.
Since stability may be hard to check directly, the inconsistency is most 
often uncovered by examining the spectrum of the putative model at \LG\ 
\DKthree. Models  whose \LG\ spectra lead to an anomaly in the discrete 
R-symmetry in spacetime can often be ``cured" by imposing additional  
fermionic gauge symmetries.

Conversely, one might easily formulate a model with ``too many" fermionic 
gauge symmetries. In that case, one might find that, at a generic point 
in the K\"ahler moduli space, it is impossible to find a choice of $E$'s 
so that the matrix $M$ is everywhere nondegenerate on the space of zeroes of 
the scalar potential.
In that case, one finds that some $\sigma$'s become noncompact in 
codimension {\it zero}. That is, the {\it generic} point in the toric 
moduli space gives rise to a singular (0,2) model.

This singularity is often easiest to see in a phase in which $p$ is
nonzero. Then,
$G$ and all the $F$'s must vanish in the vacuum as well.
We would now like to
consider if all the $E$'s for {\it any} fermionic gauge symmetry might also
vanish.
If this happens, then all the $\sigma$'s would become noncompact, and we would
 have a singularity in the linear $\sigma$-model. The main strategy
to establish that some examples exist, therefore,
is to show that the degrees of $F$'s, $G$'s and the
$E$'s corresponding to an arbitrary fermionic gauge symmetry
are not generic enough to prevent them from vanishing
{\it in} the ambient toric variety. We illustrate this in a specific example:

Let $M$ be the complete intersection of a quartic and
a quintic in $\WCP{5}{1,1,1,2,2,2}$. The data for a $(0,2)$ model based
on $M$ is as follows:
$$
\hbox{
\vbox{\offinterlineskip \tabskip=0pt
\halign{
#&
\vrule height 10pt depth 5pt
\enskip\hfil$#$\hfil\enskip\vrule &
\enskip\hfil$#$\hfil\enskip\vrule &
\enskip\hfil$#$\hfil\enskip\vrule &
\enskip\hfil$#$\hfil\enskip\vrule &
\enskip\hfil$#$\hfil\enskip\vrule &
\enskip\hfil$#$\hfil\enskip\vrule &
\enskip\hfil$#$\hfil\enskip\vrule &
\enskip\hfil$#$\hfil\enskip\vrule \cr\tablerule&
\phi_1&
\phi_2&
\phi_3&
\phi_4&
\phi_5&
\phi_6&
\chi&
p\cr\tablerule&
2&2&2&1&1&1&0&-8\cr\tablerule&
1&1&1&0&0&0&1&-3\cr\tablerule
}}
\qquad
\qquad
\vbox{\offinterlineskip \tabskip=0pt
\halign{
#&
\vrule height 10pt depth 5pt
\enskip\hfil$#$\hfil\enskip\vrule &
\enskip\hfil$#$\hfil\enskip\vrule &
\enskip\hfil$#$\hfil\enskip\vrule &
\enskip\hfil$#$\hfil\enskip\vrule &
\enskip\hfil$#$\hfil\enskip\vrule &
\enskip\hfil$#$\hfil\enskip\vrule \cr\tablerule&
\Lambda_1&
\Lambda_2&
\Lambda_3&
\Lambda_4&
\Gamma_1&
\Gamma_2\cr\tablerule&
0&1&1&6&-4&-5\cr\tablerule&
1&0&0&2&-2&-2\cr\tablerule
}}}
$$

$$
\hbox{
\vbox{\offinterlineskip \tabskip=0pt
\halign{
#&
\vrule height 10pt depth 5pt
\enskip\hfil$#$\hfil\enskip\vrule &
\enskip\hfil$#$\hfil\enskip\vrule &
\enskip\hfil$#$\hfil\enskip\vrule &
\enskip\hfil$#$\hfil\enskip\vrule &
\enskip\hfil$#$\hfil\enskip\vrule &
\enskip\hfil$#$\hfil\enskip\vrule &
\enskip\hfil$#$\hfil\enskip\vrule &
\enskip\hfil$#$\hfil\enskip\vrule &
\enskip\hfil$#$\hfil\enskip\vrule &
\enskip\hfil$#$\hfil\enskip\vrule &
\enskip\hfil$#$\hfil\enskip\vrule &
\enskip\hfil$#$\hfil\enskip\vrule \cr\tablerule&
E'_1&
E'_2&
E_1&
E_2&
E_3&
E_4&
F_1&
F_2&
F_3&
F_4&
G_1&
G_2
\cr\tablerule&
4&3&0&1&1&6&8&7&7&2&4&5\cr\tablerule&
1&1&1&0&0&2&2&3&3&1&2&2\cr\tablerule
}}
}
$$

For convenience we have indicated the degrees of the various polynomials
required. $pE'_1$ and $pE'_2$ correspond to the functions used in fermionic
gauge transformations of $\Gamma_1$ and $\Gamma_2$ respectively.
There are 2 hybrid phases in the $(0,2)$ phase diagram of this model,
corresponding respectively to $\phi_1, \phi_2, \phi_3, \chi$ or
$\phi_4, \phi_5, \phi_6$ not simultaneously vanishing in the toric
variety.
Our aim is to show that in each phase one cannot choose polynomials specified
in the bottom table that simultaneously vanish only outside the ambient
space. Let us
first consider the phase where $\phi_1, \phi_2, \phi_3$ and $\chi$ cannot be
all zero. Here, because of transversality in the ordinary 
Calabi-Yau phase ($r_1 > 0$, $r_1 - 2 r_2 > 0$),
$\phi_4 = \phi_5 = \phi_6 = 0$. It can be seen
{}from the $D$-terms that the remaining bosonic fields
form a $\WCP{3}{1,1,1,4}$. This means that,
among the $E$'s, $F$'s and $W$'s, the only ones that are nonvanishing 
are those that can be written only in terms of
$\phi_1, \phi_2, \phi_3$ and $\chi$. It is clear from the tables above that
such a polynomial must have
multi-degree $(m,n)$ where $m \le 2n$. This leaves only $G_1$, $F_4$ and $E_1$,
which are sufficient only to cut $\WCP{3}{1,1,1,4}$ to a single point at best.
The proof for the other phase is even simpler. Here similarly we have a
$\BP^2$ of points where $\phi_1, \phi_2, \phi_3$ and $\chi$ simultaneously
vanish. This means that functions which are expressible in terms of only
$\phi_4, \phi_5$ and $\phi_6$ will not vanish, which leaves only $E_2$ and
$E_3$, which again at best cuts $\BP^2$ down to a point.
Thus we see in this case that even if we can actually construct $E$'s
and $F$'s which satisfy \esuperinv, they will simultaneously vanish,
 and hence lead to noncompact $\sigma$'s, at some loci in the hybrid vacuum.

It appears that this is a model which is generically singular
at the level of perturbative string theory.  Perhaps some nonperturbative 
string effects might cure the disease, but 
in this paper, we will be content with understanding models that
are closely related to these, but avoid the subtlety of possibly
having an important dependence on higher order quantum effects.
The example
just discussed serves to illustrate our approach.
The point is
that, in this example, {\it no choices} of $G$'s or $F$'s will give an actual
\LG\ phase anywhere in the K\"ahler moduli space. This again
follows from a similar degree-matching argument as above, except now only
with the $F$'s and $G$'s. In the phase where $\phi_1, \phi_2, \phi_3$ and
$\chi$ cannot simultaneously vanish, we can at most cut the $\WCP{3}{1,1,1,4}$
with $G_1$ and $F_4$ down to a curve. Similarly for the other phase,
where all of the $G$'s and $F$'s must vanish on the $\BP^2$ given above.

This suggests that whereas one naively
would have assumed a fermionic gauge symmetry was necessary
due to $\Lambda_1$ being uncharged under the first $U(1)$, it isn't actually
the case. Nowhere in the phase diagram is there a \LG\ phase
where the gauge group is actually
broken to leave a discrete subgroup of the first
$U(1)$, as would be typical in $(2,2)$ models.
What one finds instead are two hybrid phases,
the transition from one to the other involving a {\it gauged} \LG\ theory.
In such a theory one has an unbroken $U(1)$ even though the
vacuum is given by the typical \LG\ form of fixed $p$ and
vanishing $\phi$'s and $\chi$. Further studies of the conformal
field theory of hybrids or gauged \LG\ vacua may confirm that the
model as defined by the data above gives rise to a sensible 
spectrum. Here we will assume that is the case
for some of the examples to be discussed later.

In summary, we consider only those models for which the fermionic gauge 
symmetries are such as to allow the model to be generically 
nonsingular on the full toric moduli space.

\subsec{Coherent Sheaves and $(0,2)$ models}
	In section 1 we saw that, if a model is sufficiently well behaved,
the massless left moving fermions would transform as sections of the bundle
defined by \eSeqtzf, \eSeqtzs\ over a Calabi-Yau base.
If the first map in \eSeqtzf\ is not
one-to-one, then we need to modify it by the kernel as follows:
\eqn\eSeqtzfm{ 0 \longra \oplus^{l} \cO_K \longra \oplus^{l} \cO
{\buildrel {\otimes E^a_j} \over \longra}
\oplus_{a=1}^{m} \cO(\qt_a^{(1)}, \qt_a^{(2)}, ... , \qt_a^{(s)})
 {\longra} \varepsilon \longra 0}
where $K$ is the locus on which  the map given by  the $E$'s fails to be
one-to-one. This could occur, for instance, if the $E^a_j$'s for fixed $j$
simultaneously vanish at some loci in the vacuum. In an ordinary
Calabi-Yau phase where $p$ vanishes, this would imply that the corresponding
$\sigma_j$ becomes noncompact, a signal of a singularity in the infrared.

	Similarly, if the $F$'s simultaneously vanish at some loci $S$ in the
vacuum, then \eSeqtzs\ is no longer exact as the last map is no longer
surjective. An exact sequence can be obtained by adding the cokernel of the
last map as follows:
\eqn\eSeqtzsm{ 0 \longra V \longra {{\cal E}}
{\buildrel {\otimes F_a} \over \longra}
\cO(-q_0^{(1)}, -q_0^{(2)}, ... , -q_0^{(s)}) \longra \cO_S(-q_0^{(1)},
-q_0^{(2)}, ... , -q_0^{(s)}) \longra 0 }
where $\cO_S(-q_0^{(1)}, ... , -q_0^{(s)})$ is restricted to $S$. $V$ is now
generally not a vector bundle but rather a coherent sheaf. As discussed
previously, in a Calabi-Yau phase, if $p$ nonetheless
remains compact --- spanning suitable {\it protuberances} over $M$ ---
string theory is expected to be well  behaved.

 We should hence think of $(0,2)$ models as sheaves
rather than $V$-bundles defined over Calabi-Yau spaces. In the
next section we show how this is reflected in the complex structure
variations of some models.

\subsec{Complex structure variations in $(0,2)$ models}
	Consider what happens in a $(0,2)$ model over $M$ defined by
$G = 0$ as we vary the complex structure of $G$. Unlike in the
$(2,2)$ case, the bosonic potential \ebosonicp\ does not generally depend
on the derivatives of $G$. In $(2,2)$ models,
these derivatives are  crucial in establishing that $p$ is
compact in a typical Calabi-Yau
phase. A singular $G$ in a $(2,2)$ model would therefore not make
perturbative sense.
Here, however, the $F^a$'s replace the derivatives of $G$. It is hence
possible that the linear $\sigma$-model is largely insensitive to changes in
$M$'s complex structure. We now pursue this issue in a clear-cut example.

	Consider the example studied in page 19 of \rDGM,
which we summarize by the data in the following table:
$$
\hbox{
\vbox{\offinterlineskip \tabskip=0pt
\halign{
#&
\vrule height 10pt depth 5pt
\enskip\hfil$#$\hfil\enskip\vrule &
\enskip\hfil$#$\hfil\enskip\vrule &
\enskip\hfil$#$\hfil\enskip\vrule &
\enskip\hfil$#$\hfil\enskip\vrule &
\enskip\hfil$#$\hfil\enskip\vrule &
\enskip\hfil$#$\hfil\enskip\vrule &
\enskip\hfil$#$\hfil\enskip\vrule \cr\tablerule&
\phi_1&
\phi_2&
\phi_3&
\phi_4&
\phi_5&
\chi&
p\cr\tablerule&
2&2&2&1&1&0&-9\cr\tablerule&
1&1&1&0&0&1&-5\cr\tablerule
}}
\qquad
\qquad
\vbox{\offinterlineskip \tabskip=0pt
\halign{
#&
\vrule height 10pt depth 5pt
\enskip\hfil$#$\hfil\enskip\vrule &
\enskip\hfil$#$\hfil\enskip\vrule &
\enskip\hfil$#$\hfil\enskip\vrule &
\enskip\hfil$#$\hfil\enskip\vrule &
\enskip\hfil$#$\hfil\enskip\vrule \cr\tablerule&
\Lambda_1&
\Lambda_2&
\Lambda_3&
\Lambda_4&
\Gamma\cr\tablerule&
1&1&2&5&-8\cr\tablerule&
0&0&2&3&-4\cr\tablerule
}}}
$$

Because no fermionic gauge symmetries are necessary here,
we do not need to worry about $E$'s in this model.
We now make the following choices for the polynomials defining
the manifold and the gauge bundle:
\eqn\efourdat{\eqalign{G &= H(\phi_1, \phi_2, \phi_3, \phi_4, \phi_5, \chi) \cr
	F_1 &= \phi_1^4 \chi\cr
	F_2 &= \phi_2^4 \chi \cr
	F_3 &= \phi_4^7 \chi^3 \cr
	F_4 &= \phi_3^2}}

In the above $H$ is an arbitrary polynomial of multidegree $(8,4)$ which
contains a nonzero term in $\phi_5^8 \chi^4$. These
choices ensure that the bundle is transverse in all phases except
one. In the latter phase, as alluded to previously,
 $p$ nonetheless remains compact over the base. By varying $H$
arbitrarily over the relevant space of polynomials, one
encounters various singularities in the base manifold which
are {\it irrelevant} to the linear $\sigma$-model.
The left-moving fermions here
transform as sections of a coherent sheaf $V$ while the
right-moving fermions transform as sections of the appropriate
tangent sheaf $T$.

\newsec{Aspects of $(0,2)$ Moduli Space}

An important aspect of $(2,2)$ moduli space is that sometimes there
isn't a unique desingularization of a Calabi-Yau base. In
the $(2,2)$ toric formalism this arises by different triangulations
of the toric point set data; correspondingly, in the linear sigma
model it arises by different classes of choices for the Fayet-Iliopoulos
coefficients. Physically, this gave rise to the first kind
of smooth topology changing transitions in string theory. These $(2,2)$
transitions have an immediate $(0,2)$ extension since the bundle part
of our desingularization formalism is insensitive to distinct
triangulations of the toric base. So, for instance, the
$(2,2)$ model with $h^{1,1}= 2$ given by a Calabi-Yau hypersurface in
$\WCP{4}{3,2,2,1,1}$ has the phase diagram given in Figure 1.

In figure 1 region II corresponds to an ordinary Calabi-Yau phase.
Region I corresponds to a Calabi-Yau phase related to II by a
flop. Region III corresponds to a Calabi-Yau orbifold
with $\BZ_2$ quotient singularities. Region IV is a \LG\ phase
and region V a hybrid
phase. In the geometrical phases the left and right moving fermions both
transform as sections of the tangent bundle.

\iffigs
\topinsert
$$\vbox{\centerline{\epsfysize=3.5in\epsfbox{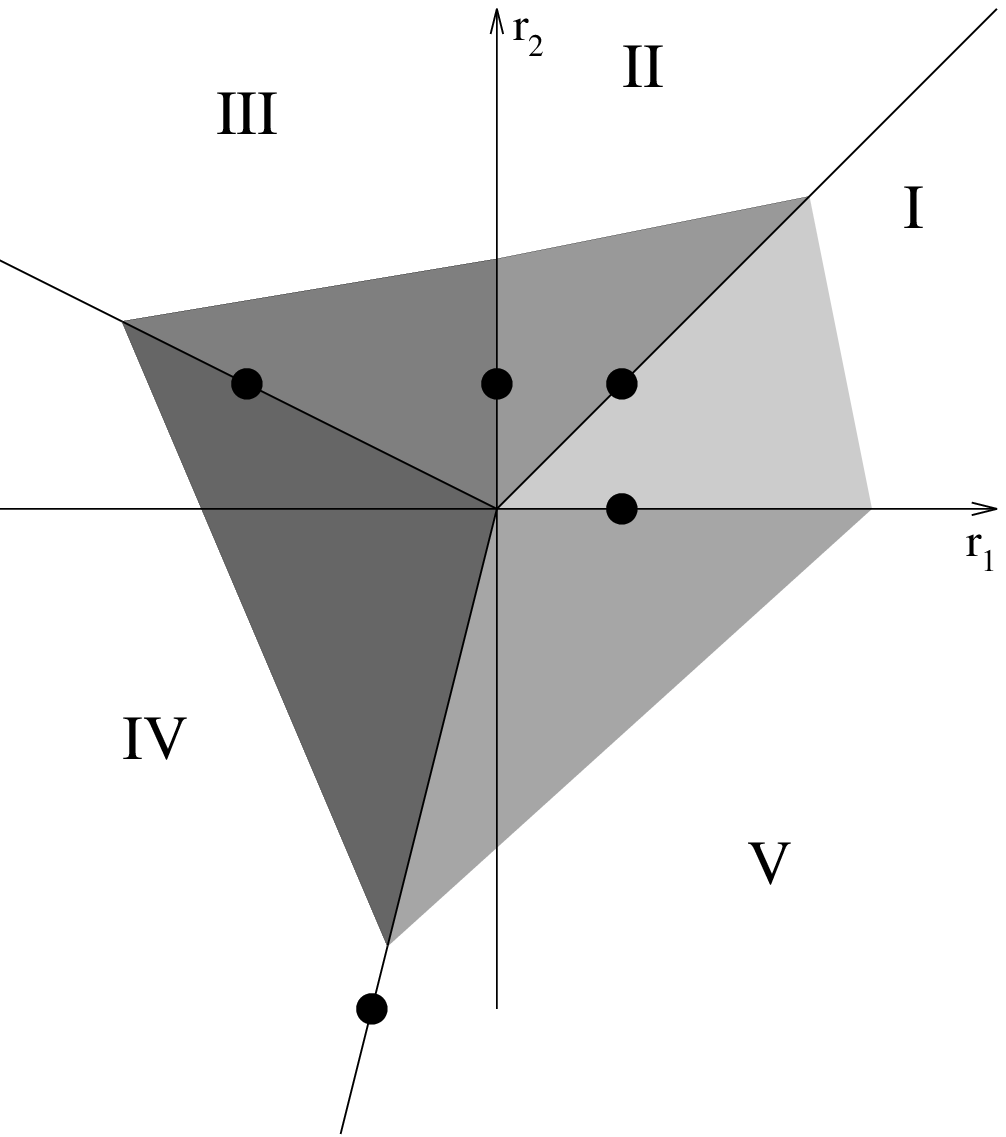}}
\centerline{Figure 1. Phase diagram for a Calabi-Yau in $\WCP{4}{3,2,2,1,1}$}
}$$
\endinsert
\fi

We can now build a $(0,2)$ model on this base whose initial
bundle data over the singular Calabi-Yau hypersurface is given by
$$
\hbox{
\vbox{\offinterlineskip \tabskip=0pt
\halign{
#&
\vrule height 10pt depth 5pt
\enskip\hfil$#$\hfil\enskip\vrule &
\enskip\hfil$#$\hfil\enskip\vrule &
\enskip\hfil$#$\hfil\enskip\vrule &
\enskip\hfil$#$\hfil\enskip\vrule &
\enskip\hfil$#$\hfil\enskip\vrule &
\enskip\hfil$#$\hfil\enskip\vrule \cr\tablerule&
\phi_1&
\phi_2&
\phi_3&
\phi_4&
\phi_5&
p\cr\tablerule&
3&2&2&1&1&-9\cr\tablerule
}}
\qquad
\qquad
\vbox{\offinterlineskip \tabskip=0pt
\halign{
#&
\vrule height 10pt depth 5pt
\enskip\hfil$#$\hfil\enskip\vrule &
\enskip\hfil$#$\hfil\enskip\vrule &
\enskip\hfil$#$\hfil\enskip\vrule &
\enskip\hfil$#$\hfil\enskip\vrule &
\enskip\hfil$#$\hfil\enskip\vrule &
\enskip\hfil$#$\hfil\enskip\vrule \cr\tablerule&
\Lambda_1&
\Lambda_2&
\Lambda_3&
\Lambda_4&
\Lambda_5&
\Gamma\cr\tablerule&
1&1&2&2&3&-9\cr\tablerule
}}}
$$

This can be desingularized by

$$\hbox{
\vbox{\offinterlineskip \tabskip=0pt
\halign{
#&
\vrule height 10pt depth 5pt
\enskip\hfil$#$\hfil\enskip\vrule &
\enskip\hfil$#$\hfil\enskip\vrule &
\enskip\hfil$#$\hfil\enskip\vrule &
\enskip\hfil$#$\hfil\enskip\vrule &
\enskip\hfil$#$\hfil\enskip\vrule &
\enskip\hfil$#$\hfil\enskip\vrule &
\enskip\hfil$#$\hfil\enskip\vrule \cr\tablerule&
\phi_1&
\phi_2&
\phi_3&
\phi_4&
\phi_5&
\chi&
p\cr\tablerule&
1&0&0&1&1&-2&-1\cr\tablerule&
1&1&1&0&0&1&-4\cr\tablerule
}}
\qquad
\qquad
\vbox{\offinterlineskip \tabskip=0pt
\halign{
#&
\vrule height 10pt depth 5pt
\enskip\hfil$#$\hfil\enskip\vrule &
\enskip\hfil$#$\hfil\enskip\vrule &
\enskip\hfil$#$\hfil\enskip\vrule &
\enskip\hfil$#$\hfil\enskip\vrule &
\enskip\hfil$#$\hfil\enskip\vrule &
\enskip\hfil$#$\hfil\enskip\vrule \cr\tablerule&
\Lambda_1&
\Lambda_2&
\Lambda_3&
\Lambda_4&
\Lambda_5&
\Gamma\cr\tablerule&
-1&-1&2&0&1&-1\cr\tablerule&
1&1&0&1&1&-4\cr\tablerule
}}}$$

In the above, we have changed basis so that the original first $U(1)$
charge is given by $Q_1 + 2 Q_2$ where the $Q_i$'s refer to the $i^{th}$
current $U(1)$ charges. An appropriate set of equations for the manifold and
gauge bundle are as follows:
\eqn\efivedat{\eqalign{G &= \phi_1 (\phi_2^3 + \phi_3^3) + (\phi_4^9 +
\phi_5^9)
\chi^4
\cr
	F_1 &= \phi_2^3 \phi_4^2 \cr
	F_2 &= \phi_3^3 \phi_5^2 \cr
	F_3 &= (\phi_4^7 + \phi_5^7) \chi^4 \cr
	F_4 &= \phi_3^3 \phi_4 + \phi_2^3 \phi_5 + \phi_1 \phi_2 \phi_3 +
\phi_4^4 \phi_5^3 \chi^3 \cr
	F_5 &= \phi_1^2 \chi}}

The above equations ensure bundle transversality in the ordinary
Calabi-Yau phase. If this theory had a \LG\ phase, then the fact that 
$\Gamma$ has
the same charge as $p$ would certainly require a fermionic
gauge symmetry and hence specification of the relevant $E$'s. 
In this case, however, the charges are 
such that 
there is no ordinary \LG\ phase in the 
K\"ahler moduli space, for any choice of manifold and bundle data. 
Hence, adopting the philosophy of section 3.2, 
we assume that the above model is defined {\it without} any fermionic gauge 
symmetries.

The phase diagram is again the same as Figure 1, although the
geometric
interpretation of each phase, besides involving the gauge bundle is
also somewhat different 
in the
$(0,2)$ case. Region II corresponds as expected to a vector bundle
over a smooth Calabi-Yau. Regions I and III, however, are now
coherent sheaves as the polynomials above do vanish simultaneously
in the ambient toric variety. Regions IV and V are their appropriate
desingularizations, which turn out, in this case, to be
``extended hybrids''. This relatively unfamiliar form of
hybridization has the expectation value of $p$ being compact
but not {\it fixed} in the semiclassical vacuum. Nevertheless,
some of the bosons
transform as a point in a $V$-bundle over the compact space
formed by the rest, a distinct feature of typical hybrid vacua.

Notice that in going from region II to region I we alter the
topology of the Calabi-Yau by flopping
a curve. The bundle data and the physical model make perfectly
good sense on the flopped model, and hence we directly see
the immediate extension of flop transitions to this more general setting.

\subsec{Multiple $(0,2)$ Resolutions}
The additional freedom in $(0,2)$ models, though, gives rise to new features
not present in the $(2,2)$ case. One of these is related to the above,
so we describe it first. In each (geometric) phase of a $(2,2)$ model,
the structure of the tangent bundle is uniquely inherited from
the structure of the base. The latter, in the toric formalism,
is determined by the charges of the chiral fields. In
a given geometrical phase of the base for a $(0,2)$ model, however, there can
be {\it different sets of charges} which desingularize the initial $V$-bundle.
In general these different charges correspond to topologically distinct total
spaces
of the pull-back of the bundle over the desingularized manifold.
By passing through the geometrically singular locus of the original
unresolved model, physics can apparently pass from one such $(0,2)$
model to another. We note that the number of generations will
not change in such a transition, and therefore these transitions are
somewhat akin to flop transitions. Here, though, the charges of the linear
sigma model fields --- not just the value of Fayet-Iliopoulos parameters ---
actually change, all in a perturbatively smooth manner.

As an example of the multiple resolutions, the following provides a valid
data for a gauge bundle over a singular Calabi-Yau hypersurface $M$
in $\WCP{4}{7,2,2,2,1}$:
$$
\hbox{
\vbox{\offinterlineskip \tabskip=0pt
\halign{
#&
\vrule height 10pt depth 5pt
\enskip\hfil$#$\hfil\enskip\vrule &
\enskip\hfil$#$\hfil\enskip\vrule &
\enskip\hfil$#$\hfil\enskip\vrule &
\enskip\hfil$#$\hfil\enskip\vrule &
\enskip\hfil$#$\hfil\enskip\vrule &
\enskip\hfil$#$\hfil\enskip\vrule \cr\tablerule&
\phi_1&
\phi_2&
\phi_3&
\phi_4&
\phi_5&
p\cr\tablerule&
7&2&2&2&1&-13\cr\tablerule
}}
\qquad
\qquad
\vbox{\offinterlineskip \tabskip=0pt
\halign{
#&
\vrule height 10pt depth 5pt
\enskip\hfil$#$\hfil\enskip\vrule &
\enskip\hfil$#$\hfil\enskip\vrule &
\enskip\hfil$#$\hfil\enskip\vrule &
\enskip\hfil$#$\hfil\enskip\vrule &
\enskip\hfil$#$\hfil\enskip\vrule &
\enskip\hfil$#$\hfil\enskip\vrule &
\enskip\hfil$#$\hfil\enskip\vrule \cr\tablerule&
\Lambda_1&
\Lambda_2&
\Lambda_3&
\Lambda_4&
\Lambda_5&
\Lambda_6&
\Gamma\cr\tablerule&
4&3&2&2&1&1&-14\cr\tablerule
}}}
$$

On resolving the base manifold, both of the following are valid
solutions to the anomaly cancellation
condition for the gauge bundle:
$$
\hbox{
\vbox{\offinterlineskip \tabskip=0pt
\halign{

\vphantom{\vrule height 10pt depth 5pt}
\enskip\hfil$#$\hfil\enskip &
\enskip\hfil$#$\hfil\enskip &
\enskip\hfil$#$\hfil\enskip &
\enskip\hfil$#$\hfil\enskip &
\enskip\hfil$#$\hfil\enskip &
\enskip\hfil$#$\hfil\enskip\cr
\Lambda_1&
\Lambda_2&
\Lambda_3&
\Lambda_4&
\Lambda_5&
\Lambda_6
\cr\tablerule
4&3&2&2&1&1\cr
2&2&2&1&0&0\cr\tablerule
4&3&2&2&1&1\cr
3&1&1&1&1&0\cr\tablerule
}}}$$

\newsec{$(0,2)$ Duality}

	Part of the
motivation for the present work is the observation of
a novel kind of duality found in \rDK. In that paper it was found
that apparently distinct geometrical $(0,2)$ models can become
isomorphic in their \LG\ phases. This opens up the
remarkable possibility of perturbative transitions from
a $(0,2)$ model on $(M_1,V_1)$ to one on $(M_2,V_2)$, with
subscripts denoting topologically distinct data. As emphasized
in \rDK, though, their analysis was severely hampered by
only  working on one-dimensional singular loci in the respective
moduli spaces. This prevented direct comparison between topological
properties in \LG\ and geometrical phases, as the latter
were not desingularized. It also left open the possibility that
subtle differences in the respective desingularizations of
the two geometrical models might lead to previously undetected
differences in their \LG\ realizations, thus spoiling
the identification. Finally, working only on the singular locus
made it impossible to extract any global information regarding how
the relevant moduli spaces might attach. The desingularizing
work reviewed above allows us to at least partially address some
of these issues, as we now do.

\subsec{Duality for complete intersections}
Let us first recall the set up in \rDK. We consider a complete-intersection
Calabi-Yau $M_1$ in a product of weighted projective spaces. A $(0,2)$
model can be built on the base as a coherent sheaf $V_1$ over $M_1$.
We now consider a phase in the K\"ahler moduli space in which
$|p|$ is forced to be nonzero in the semiclassical vacuum. This occurs for
instance in a typical hybrid phase. By making an appropriate choice of $F$'s,
one may ensure the nonzero value of $p$ is constant and proportional to
a certain linear combination of the Fayet-Iliopoulos parameters. 
Taking that combination to infinity
so as to fully freeze any
point-like anti-instantons contributed by $p$ in the $\sigma$-model,
we may then regard $p$ as a constant $<p>$. The resulting superpotential
has the form
\eqn\eSPd{\int d^2 z d\theta (\Gamma^b G_b(\Phi_i) + \Lambda^a <p>
F_a(\Phi_i))}

	We now  consider
a change of variables defined by tilded quantities as follows:
$$\left.\eqalign{\td{\Gamma}_1 &= <p> \Lambda_1\cr \td{G}_1 &=
F_1\cr
\td{E}^{'(j)}_1 &= E^{(j)}_1} \right.
\left.\eqalign{\vphone
\td{\Gamma}_2 &= <p> \Lambda_2 \cr \td{G}_2 &= F_2\cr
\vphtwo \td{E}^{'(j)}_2 &= E^{(j)}_2}\right.
\left.\eqalign{\td{\Lambda}_1 &= {\Gamma_1 \over <p>} \cr \td{F}_1 &= G_1 \cr
\vphtwo \td{E}^{(j)}_1 &= E^{'(j)}_1} \right.
\left.\eqalign{\td{\Lambda}_2 &= {\Gamma_2 \over <p>} \cr \td{F}_2 &= G_2 \cr
\vphtwo \td{E}^{(j)}_2 &= E^{'(j)}_2}\right. \eqnn\edexch \eqno{\edexch}$$
The $(0,2)$ action has the same form in terms of
the new quantities as it had in terms of the old
ones, i.e. \edexch\ is an automorphism of the theory.
Now, by taking the large radius limit again would result, if the
exchange was done in a way consistent with anomaly cancellation, in a
$(0,2)$ model over a distinct complete-intersection Calabi-Yau!
This is clear from \edexch, where the new $G$'s are related to the old $F$'s.
Similarly, the new gauge sheaf is defined by $G_1$ and $G_2$, instead of
$F_1$ and $F_2$, and the remaining $F$'s. We
can thus view this effectively as an exchange of complex structure and gauge
 bundle moduli.
Notice that the bosonic phase diagram remains identical, because the charges
of the right-moving bosons remain the same, but that, a priori, the
new and the old Calabi-Yaus are not related.

The duality noted essentially involves a permutation of the
left-moving fermionic superfields
when certain bosonic fields are frozen in moduli space.
After desingularization, we can 
imagine a generalization of the previous discussion whereby, at some
point in the moduli space, in addition
to $p$, some of the extra chiral superfields introduced by the resolution
of the base manifold are also frozen to be constants.
For simplicity we will consider here only the case for which one additional
field $\chi$ is frozen. In this case, the $G$'s and $F$'s in \eSPd\
will be functions of the $\phi$'s and $\chi$, and generally one can perform
exchanges of the $\Gamma$'s and $\Lambda$'s treating $p$ and $\chi$ as
nonzero constants. A simple attempt to define
a $\chi$-duality therefore goes as follows:
$$\left.\eqalign{\td{\Gamma}_1 &= {<p> \Lambda_1 \over <\chi>}
\cr
\td{G}_1 &= \chi F_1\cr
\td{E}^{'(j)}_1 &= {E^{(j)}_1 \over \chi}}\right.
\left.\eqalign{\vphone
\td{\Gamma}_2 &= <p> \Lambda_2 \cr \td{G}_2 &= F_2\cr
\vphthr \td{E}^{'(j)}_2 &= E^{(j)}_2}\right.
\left.\eqalign{
\td{\Lambda}_1 &= {\Gamma_1 \over <p>} \cr \td{F}_1 &= G_1 \cr
\vphthr \td{E}^{(j)}_1 &= E^{'(j)}_1} \right.
\left.\eqalign{\td{\Lambda}_2 &= {\Gamma_2 \over <p>} \cr \td{F}_2 &= G_2 \cr
\vphthr \td{E}^{(j)}_2 &= E^{'(j)}_2}\right. \eqnn\edmexch
\eqno{\edmexch}$$
where $E_1^{(j)}$ is divisible by $\chi$. The $(0,2)$ action
is invariant under \edmexch, however the resulting model is never anomaly-free.
The simplest way to obtain an anomaly free model is to, in addition to
redefining various quantities as in \edmexch, also add another left-moving
fermion $\td{\Lambda}$ whose charge is precisely that of $\chi$. One way of
understanding how this can be done is as follows.

We can consider the original model on $M_1$ but with an extra
gauge-neutral
fermion $\Lambda$. In the superpotential, $\Lambda$ multiplies
$P F(\phi_i,
\chi)$ where $F$ is a polynomial of degree $-P$. We then extend
\edmexch\ by the following exchange:
\eqn\eextexch{\td{\Lambda} = \chi \Lambda \qquad \td{E}^{(j)} =
\chi E^{(j)}
\qquad \td{F} = {{F} \over \chi}}

The gauge-neutral fermion becomes a charged
fermion after
the exchange. This allows us to understand one way in which
gauge-neutral
fermions could come about in the first place. They are simply the relics of
performing $\chi$-duality in the opposite direction.

Before proceeding with details of some concrete examples, we
briefly
examine the consequences for singularities after performing the
transformations discussed so far.

\subsec{Singularities and Duality Transformations}
As far as singularities in $\sigma_j$'s are concerned, they need to
be dealt with appropriately in each case.
The singularities in $p$ can however be discussed more generally.
As we learned from section 3, $p$ becomes noncompact when all the $G$'s
and $F$'s simultaneously vanish in the relevant toric ambient space. 
For ordinary dual models, 
the loci where $p$ becomes noncompact 
is clearly isomorphic
to that of the original model, as we have merely
interchanged the $G$'s and $F$'s in going from one model to the other.
Hence, beginning with a model with no gauge bundle 
singularities on $M_1$, we obtain one on $M_2$ by
duality, and vice versa.

We now proceed to show some examples of such dualities.
One might first try to resolve the examples discussed in
\rDK\ but it turns out that none of them are fully amenable
to toric methods. We thus concentrate on new examples which are.

\subsec{An ordinary duality for complete intersections}
We begin with an example in a product of ordinary projective spaces

to avoid having any singularities at all in the base space.
Consider then
$$
\hbox{
\vbox{\offinterlineskip \tabskip=0pt
\halign{
#&
\vrule height 10pt depth 5pt
\enskip\hfil$#$\hfil\enskip\vrule &
\enskip\hfil$#$\hfil\enskip\vrule &
\enskip\hfil$#$\hfil\enskip\vrule &
\enskip\hfil$#$\hfil\enskip\vrule &
\enskip\hfil$#$\hfil\enskip\vrule &
\enskip\hfil$#$\hfil\enskip\vrule &
\enskip\hfil$#$\hfil\enskip\vrule &
\enskip\hfil$#$\hfil\enskip\vrule \cr\tablerule&
\phi_1&
\phi_2&
\phi_3&
\phi_4&
\phi_5&
\phi_6&
\phi_7&
p\cr\tablerule&
1&1&1&1&0&0&0&-3\cr\tablerule&
0&0&0&0&1&1&1&-2\cr\tablerule
}}
\qquad
\qquad
\vbox{\offinterlineskip \tabskip=0pt
\halign{
#&
\vrule height 10pt depth 5pt
\enskip\hfil$#$\hfil\enskip\vrule &
\enskip\hfil$#$\hfil\enskip\vrule &
\enskip\hfil$#$\hfil\enskip\vrule &
\enskip\hfil$#$\hfil\enskip\vrule &
\enskip\hfil$#$\hfil\enskip\vrule &
\enskip\hfil$#$\hfil\enskip\vrule \cr\tablerule&
\Lambda_1&
\Lambda_2&
\Lambda_3&
\Lambda_4&
\Gamma_1&
\Gamma_2\cr\tablerule&
0&0&1&2&-2&-2\cr\tablerule&
1&1&0&0&-2&-1\cr\tablerule
}}}
$$
which is a complete intersection of polynomials of bidegree $(2,2)$ and
bidegree $(2,1)$ in $\CP{3} \times \CP{2}$.
The phase diagram
consists of the expected smooth Calabi-Yau phase as well as two
hybrid phases. One cannot find, for
any choice of manifold or bundle data, an
ordinary \LG\ phase in the moduli space, so we may
as before construct the model without any fermionic
gauge symmetries.
The following set of polynomials then ensure that the
bundle
is transverse in the smooth Calabi-Yau phase:
\eqn\edualdataone{\eqalign{G_1 &= (\phi_1^2 + \phi_2^2 + \phi_3^2) \phi_7^2 + 
	(\phi_1^2 + 2 \phi_3^2 + \phi_4^2) \phi_6^2 +
	(\phi_1^2 + 3 \phi_2^2 + 2 \phi_4^2) \phi_5^2 \cr
G_2 &= (\phi_1^2 + \phi_4^2) \phi_7 + (\phi_1^2
+ \phi_2^2) \phi_6 + (\phi_1^2 + \phi_3^2) \phi_5 \cr
F_1 &= \phi_4^3 \phi_6 + \phi_2^3 \phi_5 + \phi_3^3 \phi_7 + \phi_1^3 (\phi_5
+ \phi_7) \cr
F_2 &= \phi_1^3 \phi_5 	\cr
F_3 &= \phi_2^2 \phi_6^2 \cr
F_4 &= \phi_1 \phi_7^2 + \phi_2 \phi_6^2 + \phi_3 \phi_5^2 \cr
}}

We now perform a duality on the model given by the following exchange:
$$\td{G}_1 = F_1 \qquad \td{G}_2 = F_4 \qquad \td{F}_1 = G_1 \qquad
\td{F}_4 = G_2$$
This results in the following data:
$$
\vbox{\offinterlineskip \tabskip=0pt
\halign{
#&
\vrule height 12pt depth 5pt
\enskip\hfil$#$\hfil\enskip\vrule &
\enskip\hfil$#$\hfil\enskip\vrule &
\enskip\hfil$#$\hfil\enskip\vrule &
\enskip\hfil$#$\hfil\enskip\vrule &
\enskip\hfil$#$\hfil\enskip\vrule &
\enskip\hfil$#$\hfil\enskip\vrule \cr\tablerule&
\td{\Lambda}_1&
\td{\Lambda}_2&
\td{\Lambda}_3&
\td{\Lambda}_4&
\td{\Gamma}_1&
\td{\Gamma}_2\cr\tablerule&
1&0&1&1&-3&-1\cr\tablerule&
0&1&0&1&-1&-2\cr\tablerule
}}
$$
which represents, given the polynomials above,
a smooth complete intersection of two polynomials of bidegree
$(3,1)$ and $(1,2)$ in $\CP{3} \times \CP{2}$. In both
of these models the third Chern number of the respective
gauge bundles  are the same as expected, and
have the value -168. Note that the Euler number and Hodge numbers
of the two Calabi-Yau's involved are {\it not} the same. Rather,
the Hodge numbers of  the bidegree $(2,2),(2,1)$ model are given
by $h^{1,1} = 2, h^{2,1} = 62$ 
while those of the $(3,1),(1,2)$ model are given by $h^{1,1} = 2$
and $h^{2,1} = 59$.

Interesting examples abound when we start introducing weighted
projective
spaces as ambient varieties. To see how these are constructed, we
briefly discuss the relevance of these dualities to $(0,2)$
models on
hypersurfaces.

\subsec{Complete intersection dualities on hypersurfaces}
Consider a degree 9 hypersurface $M$ in $\WCP{4}{3,3,1,1,1}$.
The data for a $(0,2)$ model on $M$ is as follows:
$$
\ifx\answ\bigans\else\hskip-.25in\fi
\hbox{
\vbox{\offinterlineskip \tabskip=0pt
\halign{
#&
\vrule height 10pt depth 5pt
\enskip\hfil$#$\hfil\enskip\vrule &
\enskip\hfil$#$\hfil\enskip\vrule &
\enskip\hfil$#$\hfil\enskip\vrule &
\enskip\hfil$#$\hfil\enskip\vrule &
\enskip\hfil$#$\hfil\enskip\vrule &
\enskip\hfil$#$\hfil\enskip\vrule &
\enskip\hfil$#$\hfil\enskip\vrule \cr\tablerule&
\phi_1&
\phi_2&
\phi_3&
\phi_4&
\phi_5&
\chi&
p\cr\tablerule&
3&3&1&1&1&0&-9\cr\tablerule&
1&1&0&0&0&1&-3\cr\tablerule
}}
\
\vbox{\offinterlineskip \tabskip=0pt
\halign{
#&
\vrule height 10pt depth 5pt
\enskip\hfil$#$\hfil\enskip\vrule &
\enskip\hfil$#$\hfil\enskip\vrule &
\enskip\hfil$#$\hfil\enskip\vrule &
\enskip\hfil$#$\hfil\enskip\vrule &
\enskip\hfil$#$\hfil\enskip\vrule &
\enskip\hfil$#$\hfil\enskip\vrule &
\enskip\hfil$#$\hfil\enskip\vrule \cr\tablerule&
\Lambda_1&
\Lambda_2&
\Lambda_3&
\Lambda_4&
\Lambda_5&
\Lambda_6&
\Gamma\cr\tablerule&
3&3&1&1&1&0&-9\cr\tablerule&
1&1&0&0&0&1&-3\cr\tablerule
}}
}
$$

This $(0,2)$ model is simply a deformation of the (2,2) model, in which the
$G$ and $F$'s are, respectively,
a transverse polynomial and its derivatives with respect to
all the bosonic fields. 

We now modify the above model in a way that again satisfies the
anomaly
cancellation conditions. Namely we add a right moving boson
$\phi_6$ with
charges $(3,1)$ and a left-moving fermion $\Gamma_2$ with
charges
$(-3,-1)$ to the theory. $\Gamma_2$ here is a fermion that multiples a
polynomial $G_2$
of multi-degree $(3,1)$ in the superpotential. We may
therefore simply choose
that polynomial to be $\phi_6$. This effectively
realizes $M$ in a trivial way as a complete intersection in
$\WCP{5}{3,3,1,1,1,3}$. To satisfy \esuperinv, we need to choose
polynomials $E^1$ and $E^2$ of multi-degree $(6,2)$ for
$\Gamma_2$ under the
two fermionic gauge symmetries. One way to do so is to modify
$G$
by adding a monomial involving $\phi_6^2$: $G_{new} = G + \phi_6^2 H$.
This does not alter the vacuum structure
because $G_2 = \phi_6$ always vanishes in the vacuum. We can then solve
for $E^1$ and $E^2$ from \esuperinv\ as the terms not involving
$\phi_6$ cancel out by construction.

Embedding the model in a weighted projective space with one higher dimension
via a hyperplane section has not changed either the manifold or the bundle at
all. In the linear sigma model language, we have added a new bosonic
superfield $\Phi_6$ and a fermionic superfield $\Gamma_2$, with a mass term
which pairs them up. The infrared physics of the linear sigma model is
completely unchanged.

Thus we have a good $(0,2)$ model on $M$ realized as
a complete intersection, with toric data as follows (after 
renaming $\phi_6$ as $\phi_1$):
\ifx\answ\bigans\else\hskip-.25in\fi
\medskip\centerline{
\vbox{\offinterlineskip \tabskip=0pt
\halign{
#&
\vrule height 10pt depth 5pt
\enskip\hfil$#$\hfil\enskip\vrule &
\enskip\hfil$#$\hfil\enskip\vrule &
\enskip\hfil$#$\hfil\enskip\vrule &
\enskip\hfil$#$\hfil\enskip\vrule &
\enskip\hfil$#$\hfil\enskip\vrule &
\enskip\hfil$#$\hfil\enskip\vrule &
\enskip\hfil$#$\hfil\enskip\vrule &
\enskip\hfil$#$\hfil\enskip\vrule \cr\tablerule&
\phi_1&
\phi_2&
\phi_3&
\phi_4&
\phi_5&
\phi_6&
\chi&
p\cr\tablerule&
3&3&3&1&1&1&0&-9\cr\tablerule&
1&1&1&0&0&0&1&-3\cr\tablerule
}}
}\medskip\centerline{
\vbox{\offinterlineskip \tabskip=0pt
\halign{
#&
\vrule height 10pt depth 5pt
\enskip\hfil$#$\hfil\enskip\vrule &
\enskip\hfil$#$\hfil\enskip\vrule &
\enskip\hfil$#$\hfil\enskip\vrule &
\enskip\hfil$#$\hfil\enskip\vrule &
\enskip\hfil$#$\hfil\enskip\vrule &
\enskip\hfil$#$\hfil\enskip\vrule &
\enskip\hfil$#$\hfil\enskip\vrule &
\enskip\hfil$#$\hfil\enskip\vrule \cr\tablerule&
\Lambda_1&
\Lambda_2&
\Lambda_3&
\Lambda_4&
\Lambda_5&
\Lambda_6&
\Gamma_1&
\Gamma_2\cr\tablerule&
3&3&1&1&1&0&-9&-3\cr\tablerule&
1&1&0&0&0&1&-3&-1\cr\tablerule
}}}
In the \LG\ phase ($r_2<0$, $r_1-3r_2<0$), $\Lambda_6$ and
$\Gamma_1$ can be gauged away using the fermionic gauge symmetries.

At the \LG\ point, we can
perform an ordinary duality:
$$\td{G}_1 = F_1 \qquad \td{G}_2 = F_2 \qquad \td{F}_1 = G_1 \qquad
\td{F}_2 = G_2$$
with the corresponding $E$s and $E'$s also exchanged.
The resulting manifold is a complete intersection of two sextics in
$\WCP{5}{3,3,3,1,1,1}$.
One of the fermions, $\td{\Lambda}_1$, is neutral under both $U(1)$'s. Hence
it can be gauged away using one of the fermionic gauges symmetries in {\it
any} of the phases of the model. This complete intersection model could thus
have been formulated {\it without} $\Lambda_1$ and with only {\it one}
fermionic gauge symmetry. 

One can
check that the new model has no singularities in $\sigma$'s
in the various phases.
{}From our earlier discussion, there are no gauge bundle
singularities either. 

To see this, we consider the $E$'s as listed below:
\ifx\answ\bigans\else\hskip-.25in\fi
\medskip\centerline{
\vbox{\offinterlineskip \tabskip=0pt
\halign{
#&
\vrule height 10pt depth 5pt
\enskip\hfil$#$\hfil\enskip\vrule &
\enskip\hfil$#$\hfil\enskip\vrule &
\enskip\hfil$#$\hfil\enskip\vrule &
\enskip\hfil$#$\hfil\enskip\vrule &
\enskip\hfil$#$\hfil\enskip\vrule &
\enskip\hfil$#$\hfil\enskip\vrule &
\enskip\hfil$#$\hfil\enskip\vrule &
\enskip\hfil$#$\hfil\enskip\vrule &
\enskip\hfil$#$\hfil\enskip\vrule \cr\tablerule&
E_1&
E_2&
E_3&
E_4&
E_5&
E_6&
E'_1&
E'_2\cr\tablerule&
3\phi_2 & 3 \phi_3 & \phi_4 & \phi_5 & \phi_6 & 0 & -9 & 9\phi_1^2\cr
\tablerule&
\phi_2 & \phi_3 & 0 & 0 & 0 & \chi & -3 & 3\phi_1^2\cr\tablerule
}}}

The general solution to \esuperinv is given by
\eqn\edualdatatwo{\eqalign{
G_1&=\phi_1^2 f_1 + \phi_2 f_2 +\phi_3 f_3 +
\chi(\phi_4f_4+\phi_5f_5+\phi_6 f_6)\cr 
G_2&=f_1\cr
F_1&=3f_2\cr
F_2&=3f_3\cr
F_3&=9\chi f_4\cr
F_4&=9\chi f_5\cr
F_5&=9\chi f_6\cr
F_6&=3(\phi_4 f_4+\phi_5 f_5+\phi_6 f_6)\cr
}}
with appropriately chosen polynomials $f_{1,\dots6}$.
One choice, which leads to a bundle on a smooth Calabi-Yau, for both the
original and dual theory is
\eqn\efdatatwo{\eqalign{
f_1&=\phi_1\cr
f_2&=\phi_2^2+{\phi_3^2 \over 2} + {\chi^2 \over 6} (\phi_4^6 +
{\phi_5^6 \over 3} + {\phi_6^6 \over 4})\cr
f_3&=\phi_2^2 + \phi_1^2 +
{\chi^2 \over 6} (\phi_4^6 + \phi_5^6 + \phi_6^6)\cr
f_4&=\phi_4^6 \chi^2\cr
f_5&=\phi_5^6 \chi^2\cr
f_6&=\phi_6^6 \chi^2\cr}}
We have thus shown a perturbatively valid transition between a Calabi-Yau
hypersurface in $\WCP{4}{3,3,1,1,1}$ , which had Hodge numbers
 $h^{1,1} = 4, h^{2,1} = 112$ and a
complete intersection
in $\WCP{5}{3,3,3,1,1,1}$, which has
Hodge numbers $h^{1,1} = 5, h^{2,1} = 77$.
In the first of these, the bundle is a deformation of
the tangent bundle; in the second, it is not.
Both of these models have third Chern number of their respective
gauge bundles being -216 with, in fact, $h^1(V)=4$, and $h^2(V)=112$.
Similar methods go through for many other
hypersurfaces. 

\subsec{Chain of dualities}
Consider the degree 12 hypersurface $M$ in $\WCP{4}{4,4,2,1,1}$. We
construct as before an embedding of $M$ as a complete intersection
of a quartic and a degree 12 polynomial in $\WCP{5}{4,4,4,2,1,1}$.

The toric data for the appropriate $(0,2)$ model is given by:
\medskip\centerline{
\vbox{\offinterlineskip \tabskip=0pt
\halign{
#&
\vrule height 10pt depth 5pt
\enskip\hfil$#$\hfil\enskip\vrule &
\enskip\hfil$#$\hfil\enskip\vrule &
\enskip\hfil$#$\hfil\enskip\vrule &
\enskip\hfil$#$\hfil\enskip\vrule &
\enskip\hfil$#$\hfil\enskip\vrule &
\enskip\hfil$#$\hfil\enskip\vrule &
\enskip\hfil$#$\hfil\enskip\vrule &
\enskip\hfil$#$\hfil\enskip\vrule &
\enskip\hfil$#$\hfil\enskip\vrule \cr\tablerule&
\phi_1&
\phi_2&
\phi_3&
\phi_4&
\phi_5&
\phi_6&
\chi_1&
\chi_2&
p\cr\tablerule&
4&4&4&2&1&1&0&0&-12\cr\tablerule&
2&2&2&1&0&0&1&0&-6\cr\tablerule&
1&1&1&0&0&0&0&1&-3\cr\tablerule
}}
}\medskip\centerline{
\vbox{\offinterlineskip \tabskip=0pt
\halign{
#&
\vrule height 10pt depth 5pt
\enskip\hfil$#$\hfil\enskip\vrule &
\enskip\hfil$#$\hfil\enskip\vrule &
\enskip\hfil$#$\hfil\enskip\vrule &
\enskip\hfil$#$\hfil\enskip\vrule &
\enskip\hfil$#$\hfil\enskip\vrule &
\enskip\hfil$#$\hfil\enskip\vrule &
\enskip\hfil$#$\hfil\enskip\vrule &
\enskip\hfil$#$\hfil\enskip\vrule &
\enskip\hfil$#$\hfil\enskip\vrule \cr\tablerule&
\Lambda_1&
\Lambda_2&
\Lambda_3&
\Lambda_4&
\Lambda_5&
\Lambda_6&
\Lambda_7&
\Gamma_1&
\Gamma_2\cr\tablerule&
4&4&2&1&1&0&0&-4&-12\cr\tablerule&
2&2&1&0&0&1&0&-2&-6\cr\tablerule&
1&1&0&0&0&0&1&-1&-3\cr\tablerule
}}}

The following is an ordinary duality of the model:
$$\td{G}_1 = F_1 \qquad \td{G}_2 = F_2 \qquad \td{F}_1 = G_1 \qquad
\td{F}_2 = G_2$$
Performing the duality results in a complete intersection of two
octics in $\WCP{5}{4,4,4,2,1,1}$. The interesting point here is
that $M$ is actually a $K3$ fibration. The relevant $K3$ can be
found by essentially slicing off the first rows in the above tables.

That is, the following is a $(0,2)$ model on a $K3$ essentially
derived from the above model on the Calabi-Yau:
$$
\hbox{
\vbox{\offinterlineskip \tabskip=0pt
\halign{
#&
\vrule height 10pt depth 5pt
\enskip\hfil$#$\hfil\enskip\vrule &
\enskip\hfil$#$\hfil\enskip\vrule &
\enskip\hfil$#$\hfil\enskip\vrule &
\enskip\hfil$#$\hfil\enskip\vrule &
\enskip\hfil$#$\hfil\enskip\vrule &
\enskip\hfil$#$\hfil\enskip\vrule &
\enskip\hfil$#$\hfil\enskip\vrule \cr\tablerule&
\phi_1&
\phi_2&
\phi_3&
\phi_4&
\phi_5&
\chi&
p\cr\tablerule&
2&2&2&1&1&0&-6\cr\tablerule&
1&1&1&0&0&1&-3\cr\tablerule
}}
\qquad
\vbox{\offinterlineskip \tabskip=0pt
\halign{
#&
\vrule height 10pt depth 5pt
\enskip\hfil$#$\hfil\enskip\vrule &
\enskip\hfil$#$\hfil\enskip\vrule &
\enskip\hfil$#$\hfil\enskip\vrule &
\enskip\hfil$#$\hfil\enskip\vrule &
\enskip\hfil$#$\hfil\enskip\vrule &
\enskip\hfil$#$\hfil\enskip\vrule &
\enskip\hfil$#$\hfil\enskip\vrule \cr\tablerule&
\Lambda_1&
\Lambda_2&
\Lambda_3&
\Lambda_4&
\Lambda_5&
\Gamma_1&
\Gamma_2\cr\tablerule&
2&2&1&1&0&-2&-6\cr\tablerule&
1&1&0&0&1&-1&-3\cr\tablerule
}}}
$$

The duality now {\it extends} to this fiber, giving a transition of
a vector bundle over a $K3$ to a different vector bundle over another $K3$.
The second Chern number of each of these bundles over $K3$ is 24
, as they must be for anomaly cancellation. Thus,
each of these is dual to $F$-theory on the degree-84 hypersurface
in \WCP{4}{42,28,12,1,1}.

It turns out that $K3$ is also elliptically fibered, with a typical
fiber being that given by slicing off the first rows in the above:
$$
\hbox{
\vbox{\offinterlineskip \tabskip=0pt
\halign{
#&
\vrule height 10pt depth 5pt
\enskip\hfil$#$\hfil\enskip\vrule &
\enskip\hfil$#$\hfil\enskip\vrule &
\enskip\hfil$#$\hfil\enskip\vrule &
\enskip\hfil$#$\hfil\enskip\vrule &
\enskip\hfil$#$\hfil\enskip\vrule \cr\tablerule&
\phi_1&
\phi_2&
\phi_3&
\phi_4&
p\cr\tablerule&
1&1&1&1&-3\cr\tablerule
}}
\qquad
\vbox{\offinterlineskip \tabskip=0pt
\halign{
#&
\vrule height 10pt depth 5pt
\enskip\hfil$#$\hfil\enskip\vrule &
\enskip\hfil$#$\hfil\enskip\vrule &
\enskip\hfil$#$\hfil\enskip\vrule &
\enskip\hfil$#$\hfil\enskip\vrule &
\enskip\hfil$#$\hfil\enskip\vrule \cr\tablerule&
\Lambda_1&
\Lambda_2&
\Lambda_3&
\Gamma_1&
\Gamma_2\cr\tablerule&
1&1&1&-1&-3\cr\tablerule
}}}
$$

\noindent Again the duality extends to the duality over the toroidal fibers.

We can also work backwards from a duality between sheaves over $K3$ to
a duality over $K3$-fibered Calabi-Yaus and so on.

\newsec{Moduli Space Structure}

An important question that arises in all of these examples
is whether the attachment loci are in fact multicritical
points, or whether the physical model forces the respective moduli
spaces to either be isomorphic or possibly to join together 
 to fill out to an `enlarged' manifold moduli space.
The latter would be analogous to the `enlarged' K\"ahler moduli
space of $(2,2)$ models while the former would be closer
to $(2,2)$ conifold-like attachments. Of course, all
of our present discussion is perturbative. This is not
a question which we have been able to fully settle,
although preliminary evidence seems to indicate
 that
the full moduli space has a multicritical structure such as that
given  seems in Figure 2, where the loci of intersection denotes
the points in the phase diagram where the exchange takes place.

\iffigs
\midinsert
$$\vbox{\centerline{\epsfysize=2in\epsfbox{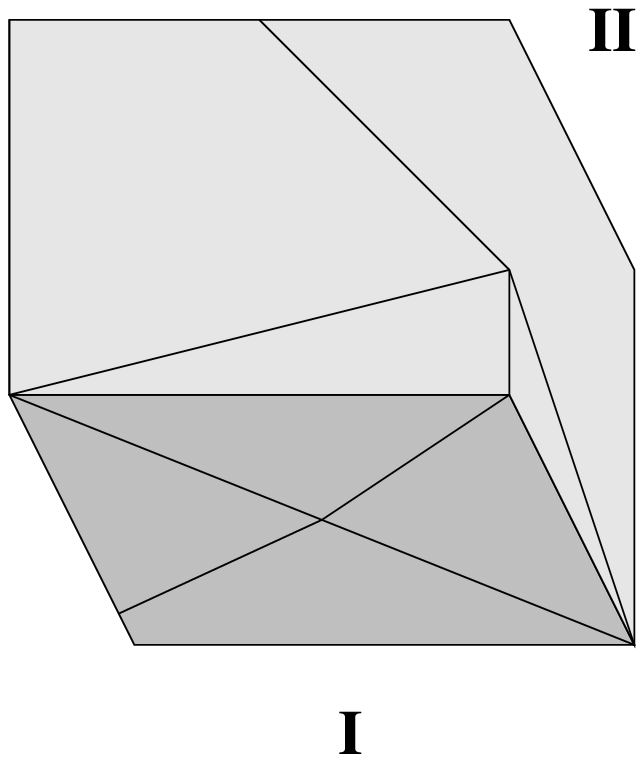}}
\centerline{Figure 2. Two moduli spaces meeting at a loci}
}$$
\endinsert
\fi

One way to settle this question, at least in specific examples, is
to calculate the bundle moduli from $H^1(EndV)$  in dual examples.
If the moduli spaces are to be identified or glued together in an
enlarged structure, we expect a one-to-one correspondence between
the full set of gauge singlet moduli (complex structure, K\"ahler structure,
holomorphic bundle structure) in each of the dual pairs.
We have made some progress in calculating $h^1(EndV)$ for some
of the models discussed here,  but the calculations are relatively
challenging.
We will not present the details here 
\nref\rCDG{work in progress.}\rCDG. 

We do note that in the example  of section {\it{5.4}}
 the moduli corresponding to 
deformations of the holomorphic structure of $V$  (in
both cases) do not all have polynomial 
representatives. That is, there are deformations of $V$ which are not captured
 by the linear sigma model. (From what we have said above, it is already 
evident that not all of the deformations of the K\"ahler structure are 
captured by the linear sigma model.) So, though we can clearly line up the 
polynomial deformations in the two theories (there are 380 of them, 
corresponding in each case to some mixture of deformations of complex 
structure and holomorphic deformations of $V$), as well as the two K\"ahler 
moduli which are captured by the linear sigma model, we do not know how to 
make a correspondence between the other moduli which are not representable in 
the linear sigma model. Indeed, as the \LG\ theory has more 
singlets (431) than are expected as moduli for either the $M$ or the 
$\tilde M$ theory, it appears likely that the \LG\ locus is a 
multicritical point, with the $M$ and $\tilde M$ theories as different 
branches of the moduli space which meet at \LG.
Fully establishing this requires completion of the geometric  calculation
of moduli, as we will present elsewhere.

If the dual models we have found are in fact multicritical,
 many distinct $(0,2)$ models
would be joined by passing through suitable points in their
enlarged K\"ahler moduli space --- points at which certain bosonic
fields are frozen to constant values. This would be a perturbative
cousin of the nonperturbative $(2,2)$ multicritical conifold transitions.

\newsec{Remarks on Toric Resolutions of $(0,2)$ Models}

Before concluding, we wanted to point out one feature of
$(0,2)$ resolutions which has not been clearly delineated
in either \rDGM\ or in the above. For a given
singular model, the number of solutions to \efcond\
and \escond\ generally
greatly exceeds the number of physically sensible resolutions.
At first sight, our discussion has identified solutions of
these equations with physical resolutions, but in reality,
equations \efcond\ and \escond\ are necessary but not sufficient
conditions. The condition that we have not addressed
is that of stability. We recall from \rWphases\ that the
bundle $V$ over a smooth base $M$ must be a stable bundle
in order to be able to solve the vanishing beta function
equations for the gauge fields. This has always been a difficult
condition to systematically incorporate into physical models,
and at present it is still unclear how to generally do so at the level
of the linear sigma model. What has been noticed \rDK, for instance,
is that certain naively sensible  choices of $V$ can sometimes lead
to pathological behaviour. The natural guess is
that this is due to failing to meet the stability condition.

In the treatment of \rDGM\ there are some additional necessary conditions
which must be satisfied beyond equations \efcond\ and \escond\ in order
for there to be a chance that the resulting solution is stable.
For instance, if $h^3(V)$ is nonzero, then $V$ is not stable
as the dual bundle $V^*$ has a global section.

These conditions turn out to be rather efficient at eliminating
solutions to
equations \efcond\ and \escond. For example, the following
are all valid solutions to the anomaly cancellation
conditions for a gauge bundle over a smooth Calabi-Yau
in $\WCP{4}{3,2,2,1,1}$:
$$
\hbox{
\vbox{\offinterlineskip \tabskip=0pt
\halign{
\vphantom{\vrule height 10pt depth 5pt}
\enskip\hfil$#$\hfil\enskip &
\enskip\hfil$#$\hfil\enskip &
\enskip\hfil$#$\hfil\enskip &
\enskip\hfil$#$\hfil\enskip &
\enskip\hfil$#$\hfil\enskip &
\enskip\hfil$#$\hfil\enskip &
\enskip\hfil$#$\hfil\enskip &
\enskip\hfil$#$\hfil\enskip\cr&
\Lambda_1&
\Lambda_2&
\Lambda_3&
\Lambda_4&
\Lambda_5&
\Lambda_6&
p
\cr\tablerule
1.&&&&2&3&5&-10\cr
&&&&0&2&3&-5\cr\tablerule
2.&&&0&1&1&15&-17\cr
&&&-1&1&1&7&-8\cr\tablerule
3.&&&0&1&1&15&-17\cr
&&&1&0&0&6&-7\cr\tablerule
4.&&1&1&2&2&3&-9\cr
&&1&1&0&1&1&-4\cr\tablerule
5.&0&1&1&2&2&3&-9\cr
&1&0&0&1&1&1&-4\cr\tablerule
}}}$$

We can see however that solution 2 is
problematic whether or not fermionic gauge symmetries are considered.
The reason is that
the relevant $E$'s for
$\Lambda_3$ have to be either $0$ or a suitable rational function.
The latter not being 
well-defined over
the ordinary Calabi-Yau phase, we must set the $E$'s
for $\Lambda_3$ to be 0. Having done so, all the $E$'s,$F$'s and
$G$'s will simultaneously
vanish in
the hybrid phases, resulting in a model that appears
to be generically
singular.
On the other hand, if we consider instead
a model defined without fermionic gauge symmetries, we find
from the exactness of \eSeqtzf\ and \eSeqtzs\ that $h^0(V^*) = 1$.
As alluded to above, this corresponds to the kernel bundle being
unstable as a perturbative solution to the string equations
of motion.

Ideally, a more robust formalism can be found in which the
necessary and sufficient conditions for a bona fide string
solution are directly applied and manifestly satisfied by
candidate solutions. As the fundamental ingredients in
the approach taken here involves line bundles over
toric varieties, it would seem possible that an extended
toric formalism should exist for this purpose. Finding such
a formalism is worthy of concerted effort. Presumably it would
also be the appropriate framework for discussing $(0,2)$ mirror
symmetry, a topic on which there has been some interesting
recent progress \rmirrortwo.

\bigbreak\bigskip\bigskip\centerline{{\bf Acknowledgements}}\nobreak
The work of TMC is supported in part by the National Science
Foundation.
The work of J.D.~is supported in part by NSF grant PHY9511632, the Robert
A.~Welch Foundation and the Alfred P. Sloan Foundation.
 The work of BRG is supported by a National Young Investigator
Award and the Alfred P. Sloan Foundation. BRG thanks the physics department 
of Rutgers University where some of this work was completed.

\listrefs

\bye